# Effect of the spin crossover of local copper-oxygen states on the electronic structure of HTSC cuprates

I. A. Makarov[1, *] and S. G. Ovchinnikov[1, †]

[1]*Kirensky Institute of Physics, Federal Research Center KSC SB RAS,*
*Akademgorodok 50, bld. 38, 660036 Krasnoyarsk, Russia*
(Dated: February 2, 2024)

In this work, the effect of uniaxial pressure along the $c$ axis on the electronic structure of the HTSC cuprate $La_{2-x}Sr_xCuO_4$ is investigated at the doping levels $x = 0.1, 0.15, 0.25$. The GTB method within the five-band p-d model framework is used to describe the electron system. The uniaxial compression leads to a significant reconstruction of the electronic structure and a change in the character of low-energy quasiparticle excitations: a large contribution of $a_{1g}$ symmetry orbitals appears at the top of the valence band. The crossover between the local Zhang-Rice singlet and the Emery-Reiter triplet was found at the pressure $P_c = 15.1$ GPa. The characteristic changes in the electronic structure under pressure occur abruptly as a result of the crossover. In particular, the top of the valence band displaces to the region around the $k$-point $(\pi, 0)$, the Fermi contour transforms to the four pockets around $(0, 0), (2\pi, 0), (0, 2\pi), (2\pi, 2\pi)$ and the one contour around $(\pi, \pi)$.

## I. INTRODUCTION

The study of the crystal, magnetic and electronic structure of systems with strong and intermediate electronic correlations using a pressure in a wide range of magnitudes has revealed a sharp change in properties caused by the change in the local magnetic moment in many compounds. The spin crossover between high-spin and low-spin states was discovered in a variety of iron oxides including orthoferrites R$FeO_3$ [1, 2] (R=La, Nd, Pr, Lu), borates $FeBO_3$ [3–5], $GdFe_3(BO_3)_4$ [6], garnet $Y_3Fe_5O_{12}$ [7], multiferroic $BiFeO_3$ [8], hematite $\alpha$-$Fe_2O_3$ [9–11] under high pressure. These compounds pass from the dielectric to the metallic state with increasing pressure according to different scenarios in which, one way or another, a spin crossover is involved. The magnetic collapse associated with the disappearance of the magnetic moments of 3d ions under pressure was predicted in the monoxides Mn-O, Fe-O, Co-O, and Ni-O [12, 13]. The magnetic moment collapse predicted in calculations [14–16] was confirmed in MnO [14], the volume collapse and the insulator-metal transition occur due to this moment collapse and simultaneously with it. The collapse of the magnetic moment is due to an increase in the energy level splitting as the crystal field grows.

In superconducting chalcogenides, the change in the properties of the magnetic structure under pressure is directly manifested in the superconducting phase, namely, the stabilization of the superconducting state occurs, the $T_c$ increases, magnetism and superconductivity coexists. It turned out that applying a pressure of 8 GPa to the compound MnP with a helimagnetic structure suppresses the magnetic order and leads to superconductivity of 1 K [17] although it was previously believed that compounds with Mn are incompatible with superconductivity due to their strong magnetism. In the studies of manganese chalcogenides MnSe and MnS [18–20], it was found that their crystal structure changes under pressure from cubic to orthorhombic with a significant (20%) reduction in the volume of the unit cell. This structural change occurs along with a change in the spin state of $Mn^{2+}$ ($d^5$) from high spin ($S = 5/2$, $t_{2g}^3 e_g^2$) to low spin ($S = 1/2$, $t_{2g}^5 e_g^0$). Later, in MnSe, a superconducting state with a critical temperature of 6.5 K at 40 GPa as well as a change from insulator to metallic behavior was discovered during the transition to the orthorhombic phase under pressure [21]. Although it is quite possible that pressure-induced superconductivity is associated with the onset of the orthorhombic phase the effect of the insulator-metal transformation and the suppression of the antiferromagnetic ordering in MnSe may play an important role.

The multiorbital nature of iron-based superconductors combined with spin and charge degrees of freedom leads to the observation of many intriguing phenomena such as magnetic or orbital ordering [22], electronic nematicity [23–25]. The unusual coexistence of magnetism and superconductivity in some iron chalcogenide compositions $Fe_{1+d}Se_{1-x}Te_x$ may result from the competition between different orbital and magnetic states. Iron chalcogenides FeSe [26–28] have a similar crystal structure and common elements of the chemical composition with $LaFeAsO_{1-x}F_y$ pnictides [29–36]. Both of these families are superconductors but their magnetic phase diagrams are significantly different. There is a long-range antiferromagnetic order [37] in undoped $LaFeAsO_{1-x}F_y$ pnictides that is destroyed with doping and only short-range magnetic correlations remain [38]. The region of existence of short-range magnetic correlations extends to the superconducting phase [38, 39] similar to what happens in cuprates. There is no long-range magnetic order in the undoped FeSe; superconductivity appears at 8 K [26, 40] against the background of the orthorhombic phase. The application of a pressure of about 9 GPa leads to an increase in the $T_c$ in the tetragonal superconducting FeSe to 37 K; this compound transforms into a hexagonal non-superconducting phase at a pressure above 9 GPa [41, 42]. Later NMR studies revealed evidence of antiferromagnetic spin fluctuations near $T_c$ under pressure [43]. The static magnetic ordering was observed above a pressure of 0.8 GPa using $\mu$SR [44]. These experiments showed that as soon as the magnetic ordering occurs, the magnetic and superconducting states appear to compete with each other. This is manifested in the fact that the incommensurate magnetic order is suppressed upon the onset of superconductivity and $T_c$ decreases in the pressure range of $0.8 \div 1.2$ GPa. Both ground states coexist above 1.2 GPa, the Neel temperature $T_N$ and $T_c$ simultaneously increases with increasing pressure and the magnetic order becomes commensurate [45].

*Electronic address: maki@iph.krasn.ru
†Electronic address: sgo@iph.krasn.ru

In undoped HTSC cuprates, strong electronic correlations (SEC) lead to the appearance of localized magnetic moments on copper atoms. The long-range magnetic order is formed due to the antiferromagnetic superexchange interaction of these moments through completely filled $p$-orbitals of oxygen atoms. The copper atoms have the $d^9$ electronic configuration; in this case, a variation of the ionic spin moment is impossible. However, it is necessary to take into account the special nature of low-energy excitations in SEC systems. In the ground state of undoped cuprates, each $Cu^{2+}$-ion has $d^9$ electrons with spin $S = 1/2$ (in hole representation single-hole state). The conduction band is formed by excitations between zero-hole and single-hole local states (upper Hubbard band, UHB), and the valence band is formed by excitations between single-hole and two-hole states (lower Hubbard band, LHB). Since the copper $3d$ states and the oxygen $2p$ states are highly hybridized in these compounds the local single-hole and two-hole states are of a copper-oxygen nature. The single-hole states correspond to mixture of electronic configurations $d^9p^6$ and $d^{10}p^5$. The two-hole states correspond to the electronic configurations $d^8p^6$, $d^{10}p^4$, $d^9p^5$. The contribution of local two-hole states will increase with hole doping. Variations of the spin momentum for two-hole states are already possible. It should be noted that the competition of local states in cuprates occurs not between pure copper states with the $d^8$ configuration, for example, $|d_{x\downarrow}d_{x\uparrow}\rangle$ and $|d_{x\downarrow}d_{z\uparrow}\rangle$, but between copper-oxygen hybridized states corresponding to the $d^8p^6$, $d^{10}p^4$, $d^9p^5$ configurations. Thus, the ground two-hole state is the Zhang-Rice singlet [46] which includes the state $|d_{x\downarrow}d_{x\uparrow}\rangle$ and not the Emery-Reiter triplet [47] including the state $|d_{x\uparrow}d_{z\uparrow}\rangle$ although the state $|d_{x\uparrow}d_{z\uparrow}\rangle$ itself is lower in energy than the state $|d_{x\downarrow}d_{x\uparrow}\rangle$ due to the Hund exchange. We will show that under uniaxial deformation it is possible to obtain the crossover between the Zhang-Rice singlet and the Emery-Reiter triplet, and not only the spin moment of the local copper-oxygen state will change but also its symmetry and electron density distribution. These changes will significantly affect the electronic structure of low-energy excitations and thus lead to unique properties.

It can be expected that a change in the local moment entailing a significant reconstruction of the electronic and magnetic structure is possible when pressure is applied in HTSC cuprates. The results of pressure experiments in cuprates revealed several main effects. Compression of the $CuO_2$ plane leads to an increase in $T_c$ [48–52] apparently due to an increase in the pairing exchange interaction when the interatomic distance Cu-O is reduced. Uniaxial compression along the $c$ axis affects $T_c$ through the two mechanisms: on the one hand, through an increase in the number of doped holes, on the other hand, through an increase in Cu-O distances in the $CuO_2$ plane [48–52]. However, abrupt changes in electronic properties of cuprates that could be a consequence of crossover of local states have not yet been detected in pressure experiments. In HTSC cuprates, the most probable spin crossover is between the Zhang-Rice singlet and the Emery-Reiter triplet since the latter is the first excited two-hole state. To make this crossover possible, it is necessary to increase the energy of a electron on $a_{1g}$ symmetry orbitals. This can be ensured by uniaxial pressure along the $c$ axis at which the tetrahedral distortion of $CuO_6$ octahedra is reduced. It is possible that the reason for the lack of evidence of a spin crossover in cuprates is the lack of experiments at sufficiently high uniaxial pressures.

Previously, a theoretical study of the effect of high uniaxial pressure on the band structure of undoped n- and p-type cuprates was carried out in [53] within the framework of the generalized tight-binding (GTB) method in the mean-field approximation. In the work [53], it was found that triplet copper-oxygen states emerge at the top of the valence band in the region of wave vectors $(0, 0)$ and $(\pi, \pi)$ at a uniaxial pressure of 220 GPa along the $c$ axis. In the work [54], the influence of the hydrostatic and uniaxial pressure on the electronic structure of doped cuprates was studied for several small pressure values. There was a slight renormalization of the orbital contributions in local cluster eigenstates, a change in the energies of local levels and in the energy interval between the local levels of different symmetries under pressure. The main effect on the electronic structure is a slight reconstruction of the dispersion and Fermi surface of quasiparticle excitations; this reconstruction manifests itself in a shift in the critical hole concentrations at which quantum phase transitions occur. The effect of the spin crossover of local copper-oxygen states on the electronic structure of HTSC cuprates has not yet been studied theoretically.

In this work, we present the theoretical study of the influence of uniaxial pressure along the $c$ axis in the doped single-layer HTSC cuprate $La_{2-x}Sr_xCuO_4$ (LSCO) on local copper-oxygen states and the electronic structure of quasiparticle excitations constructed on these local states. The magnitude of uniaxial pressure at which the crossover between local copper-oxygen states occurs is estimated, the character of these states is clarified. The effect of this crossover on the electronic structure of low-energy excitations is studied. The five-band p-d model will be used to take into account all orbitals involved in the formation of low-energy excitations. Correct accounting of SEC and the effects of strong hybridization of copper and oxygen orbitals is achieved using the GTB method [55–58]. A generalized mean-field approximation within which kinematic and spin-spin correlation functions are taken into account will be used to obtain the electronic structure.

The present paper includes six sections. Section II presents the Hamiltonian of the five-band p-d model. Section III describes the technique for describing the electronic structure of the system under consideration using the GTB method and basic information about the local many-particle states of the CuO$_6$ octahedron. Section IV is devoted to the band structure of the five quasiparticle excitations in a layer of CuO$_6$ octahedra without pressure in order to understand the features of the electronic structure constructed taking into account the local excited triplet and singlet states, and not just the ground singlet Zhang-Rice state. In Section V, the effects of uniaxial pressure along the $c$ axis on the electron system are discussed. Subsection V A presents the effects of uniaxial pressure on the local many-particle states of the CuO$_6$ octahedron. Subsection V B is devoted to the effects of uniaxial pressure on the band structure and Fermi contour of quasiparticle excitations. In Subsection V C, the electronic structure reconstruction in the crossover region is studied. Section VI presents a conclusion summarizing the main results.

## II. HAMILTONIAN OF THE FIVE-BAND P-D MODEL FOR A LAYER OF CuO$_6$ OCTAHEDRA

The real structure of LSCO in the tetragonal phase is modeled by a layer of CuO$_6$ octahedra. A single CuO$_6$ octahedron includes one copper atom in the center, four planar oxygen atoms, each of which belongs simultaneously to two unit cells, and two apical oxygen atoms. To describe the electronic structure of low-energy excitations in a layer of CuO$_6$ octahedra, we will consider five orbitals: $d_{x^2-y^2}$- and $d_{3z^2-r^2}$-orbitals of the copper atom (hereinafter will be denoted as $d_x$ and $d_z$, respectively), $p_x$- and $p_y$-orbitals of the planar oxygen atoms and the molecular bonding orbital $p_z$ formed by the two atomic $p_z$-orbitals of the apical oxygens (Fig. 1a). It is believed that the region of excitations near the Fermi level in the considered compound is well described by the orbitals $d_x$, $p_x$ and $p_y$, however, uniaxial pressure along the $c$ axis leads to an increase in the contribution of the $p_z$-orbital and the $a_{1g}$ symmetry orbitals hybridized with it to the electronic structure at the top of the valence band. The Hamiltonian of the five-band p-d model in the hole representation in the nearest neighbors approximation has the form:

$$
\begin{aligned}
H = &\sum_{\mathbf{f}\sigma} \left[ (\varepsilon_{dx} - \mu)\, n^{dx}_{\mathbf{f}\sigma} + (\varepsilon_{dz} - \mu)\, n^{dz}_{\mathbf{f}\sigma} + (\varepsilon_{pz} - \mu)\, n^{pz}_{\mathbf{f}\sigma} \right] + \sum_{\mathbf{g}\sigma} (\varepsilon_p - \mu) \left[ n^{px}_{\mathbf{g}\sigma} + n^{py}_{\mathbf{g}\sigma} \right] + \\
&\sum_{\mathbf{f}\sigma} t_{pd} \left[ d^\dagger_{x\mathbf{f}\sigma} \left( p_{x(\mathbf{f}-\mathbf{a}/2)\sigma} - p_{x(\mathbf{f}+\mathbf{a}/2)\sigma} - p_{y(\mathbf{f}-\mathbf{b}/2)\sigma} + p_{y(\mathbf{f}+\mathbf{b}/2)\sigma} \right) + h.c. \right] + \\
&\sum_{\mathbf{f}\sigma} t_{pdz} \left[ d^\dagger_{z\mathbf{f}\sigma} \left( -p_{x(\mathbf{f}-\mathbf{a}/2)\sigma} + p_{x(\mathbf{f}+\mathbf{a}/2)\sigma} - p_{y(\mathbf{f}-\mathbf{b}/2)\sigma} + p_{y(\mathbf{f}+\mathbf{b}/2)\sigma} \right) + h.c. \right] + \sum_{\mathbf{f}\sigma} t_{pzdz} \left( d^\dagger_{z\mathbf{f}\sigma} p_{z\mathbf{f}\sigma} + h.c. \right) + \\
&\sum_{\mathbf{f}\sigma} t_{pp} \left[ \left( p^\dagger_{x(\mathbf{f}-\mathbf{a}/2)\sigma} - p^\dagger_{x(\mathbf{f}+\mathbf{a}/2)\sigma} \right) \left( p_{y(\mathbf{f}-\mathbf{b}/2)\sigma} - p_{y(\mathbf{f}+\mathbf{b}/2)\sigma} \right) + h.c. \right] + \\
&\sum_{\mathbf{f}\sigma} t_{ppz} \left[ \left( p^\dagger_{x(\mathbf{f}-\mathbf{a}/2)\sigma} - p^\dagger_{x(\mathbf{f}+\mathbf{a}/2)\sigma} + p^\dagger_{x(\mathbf{f}-\mathbf{b}/2)\sigma} - p^\dagger_{x(\mathbf{f}+\mathbf{b}/2)\sigma} \right) p_{z\mathbf{f}\sigma} + h.c. \right] + \\
&\sum_{\mathbf{f}\lambda} U_d n^{d\lambda}_{\mathbf{f}\uparrow} n^{d\lambda}_{\mathbf{f}\downarrow} + \sum_{\mathbf{f}\sigma\sigma'} V_d n^{dx}_{\mathbf{f}\sigma} n^{dz}_{\mathbf{f}\sigma'} + \sum_{\mathbf{f}\sigma\sigma'} J_d d^\dagger_{x\mathbf{f}\sigma} d_{x\mathbf{f}\sigma'} d^\dagger_{z\mathbf{f}\sigma'} d_{z\mathbf{f}\sigma} \\
&\sum_{\mathbf{g}\lambda\lambda'} U_p n^{p\lambda}_{\mathbf{g}\uparrow} n^{p\lambda'}_{\mathbf{g}\downarrow} + \sum_{\langle \mathbf{fg}\rangle} \sum_{\lambda\lambda'\sigma\sigma'} V_{pp} n^{p\lambda}_{\mathbf{f}\sigma} n^{p\lambda'}_{\mathbf{g}\sigma'} + \sum_{\langle \mathbf{fg}\rangle} \sum_{\lambda\lambda'\sigma\sigma'} V_{pd} n^{d\lambda}_{\mathbf{f}\sigma} n^{p\lambda'}_{\mathbf{g}\sigma'},
\end{aligned}
\tag{1}
$$

Here $n^{d\lambda}_{\mathbf{f}\sigma} = d^\dagger_{\lambda\mathbf{f}\sigma} d_{\lambda\mathbf{f}\sigma}$, $n^{p\lambda}_{\mathbf{g}\sigma} = p^\dagger_{\lambda\mathbf{g}\sigma} p_{\lambda\mathbf{g}\sigma}$ are the operators of the number of holes for the copper and oxygen orbitals, respectively. The indices $\mathbf{f}, \mathbf{g}$ show the coordinates of the atom on which a certain orbital is located, the indices $\sigma, \sigma'$ characterize the projection of the hole spin on a cer-tain orbital, the indices $\lambda, \lambda'$ indicate the type of orbital, $\lambda = x, z$ for $d$-orbitals and $\lambda, \lambda' = x, y, z$ for $p$-orbitals. $\varepsilon_{dx}, \varepsilon_{dz}$ are on-site energies for the $d_x$-, $d_z$-orbitals of the copper atom and $\varepsilon_{p\lambda} = \varepsilon_{px}, \varepsilon_{py}, \varepsilon_{pz}$ - on-site energies for the $p_x$-, $p_y$-orbitals of the planar oxygens and the $p_z$- orbitals of the apical oxygens, respectively. The following hopping integrals are also taken into account: $t_{pd(z)}$ is the hopping integral between the copper $d_x$ ($d_z$) and $p_x$, $p_y$ planar oxygen orbitals, $t_{pzdz}$ - between the copper $d_z$ and $p_z$ apical orbitals, $t_{pp}$ - between the planar oxygen $p_x$- and $p_y$-orbitals, and $t_{ppz}$ - between the planar $p_x$, $p_y$ and $p_z$ apical orbitals. $U_d$ and $U_p$ are the parameters of the intraatomic intraorbital Coulomb interaction of holes on the copper and oxygen atoms, respectively. $V_d$ is the parameter of the interorbital intraatomic Coulomb interaction of holes on the $d_x$ and $d_z$-orbitals of the copper atom. $V_{pd}$ and $V_{pp}$ are the parameters of the interatomic Coulomb interaction of holes on the copper and oxygen atoms and on the neighboring planar oxygens, respectively. $J_d$ is the parameter of the Hund exchange interaction of holes in the $d_x$- and $d_z$-orbitals of the copper atom. The values of the Coulomb interaction parameters (in eV):

$$U_d = 9, V_d = 7, U_p = 4, V_{pd} = 1.5, V_{pp} = 1, J_d = 1 \quad (2)$$

The values of on-site energies and hopping integrals for the undeformed LSCO were obtained in [59] using *ab initio* LDA+GTB calculations. The problem of finding the dependence of all parameters $\varepsilon_{dz}, \varepsilon_p, \varepsilon_{pz}, t_{dzpz}, t_{pd}, t_{pdz}, t_{ppz}$ on pressure is divided into two subtasks, the first is to determine how the distances between atoms change under uniaxial pressure along the $c$ axis, the second is how a given change in the interatomic distance affects the energy parameters. The relationship between pressure and lattice deformation according to Hooke's law using elastic constants for the LSCO in the tetragonal phase is used to solve the first subtask. The procedure for calculating the change in the lattice parameters with pressure is described in detail in A. The change in the energy parameters of the Hamiltonian during the displacement of oxygen atoms (apical and planar) under the uniaxial pressure along the $c$ axis is described in B. The dependences of all energy parameters on uniaxial pressure obtained using the considerations of A, B are shown in Fig. 1b.

## III. THE GTB METHOD AND LOCAL MANY-PARTICLE STATES

To correctly describe electronic excitations in HTSC cuprates, it is necessary to exactly take into account SEC and covalence effects. The construction of the electronic structure of quasiparticle excitations based on exact local many-particle states of a finite-size cluster can be performed within the GTB method [55–58]. The GTB method which is a cluster form of perturbation theory in terms of Hubbard operators includes four stages. The first stage is to choose the type of a cluster and divide the crystal lattice into clusters. In this case, the Hamiltonian is represented as the sum of intracluster interactions and the sum of interactions of particles located in different clusters. The second stage is the exact diagonalization of an individual cluster to obtain its own many-particle local states. At the third stage, holes (or electrons) are represented through the sum of all possible transitions described by Hubbard operators between the eigenstates of a cluster with the number of fermions differing by one. At the fourth stage, the complete Hamiltonian is written in terms of Hubbard operators.

In this work, a $CuO_6$ octahedron is considered as a cluster. The Shastry type transformation [60] is made to orthogonalize the states of planar oxygen atoms in neighboring clusters. In this transformation, the $p_x$-, $p_y$-orbitals of planar oxygens are replaced by the molecular orbitals $b$ and $a$ in $k$-space [61]:

$$b_{\mathbf{k}} = \frac{i}{\mu_{\mathbf{k}}} \left( s_{x\mathbf{k}} p_{x\mathbf{k}} - s_{y\mathbf{k}} p_{y\mathbf{k}} \right),$$

$$a_{\mathbf{k}} = -\frac{i}{\mu_{\mathbf{k}}} \left( s_{y\mathbf{k}} p_{x\mathbf{k}} + s_{x\mathbf{k}} p_{y\mathbf{k}} \right), \quad (3)$$

where $s_{kx} = \sin(k_x a/2)$, $s_{ky} = \sin(k_y b/2)$ and $\mu_{\mathbf{k}} = \sqrt{s_{kx}^2 + s_{ky}^2}$. The complete Hamiltonian in terms of the $b$ and $a$ orbitals is given in [61]. The zero-, single-, and two-hole cluster eigenstates are obtained as a result of the exact diagonalization of the Hamiltonian of a $CuO_6$ octahedron with the corresponding number of holes. The zero-hole state is the vacuum hole state or, in other words, the completely filled electronic state $d^{10}p^6$. In the single-hole sector, the ground state is the doublet $b_1$ formed by the copper $d_x$- and oxygen $b$-orbitals. The first excited state consisting of the $d_z$-, $a$- and $p_z$-orbitals is located above the ground state by 1.8 eV at zero pressure. The ground two-hole state is the Zhang-Rice singlet

$$\left| A_1^S \right\rangle = L_{dxb} \left( |d_{x\downarrow}b_\uparrow\rangle - |d_{x\uparrow}b_\downarrow\rangle \right) + \\ L_{dxdx} |d_{x\downarrow}d_{x\uparrow}\rangle + L_{bb} |b_\downarrow b_\uparrow\rangle \quad (4)$$

(Fig. 2,left panel, black line). The first excited two-hole state is the Emery-Reiter triplet $\left| B_1^T \right\rangle = \left\{ \left| B_1^{T0} \right\rangle, \left| B_1^{T1} \right\rangle, \left| B_1^{T-1} \right\rangle \right\}$ (Fig. 2, left panel, red line) which is a mixture of the products of $b_{1g}$ and $a_{1g}$ symmetry orbitals:

$$\left| B_1^{T0} \right\rangle = L_{dxdz} \left( |d_{x\downarrow}d_{z\uparrow}\rangle + |d_{x\uparrow}d_{z\downarrow}\rangle \right) + L_{dxa} \left( |d_{x\downarrow}a_\uparrow\rangle + |d_{x\uparrow}a_\downarrow\rangle \right) + L_{dxpz} \left( |d_{x\downarrow}p_{z\uparrow}\rangle + |d_{x\uparrow}p_{z\downarrow}\rangle \right) + ..., \quad (5)$$

$$\left| B_1^{T1} \right\rangle = L'_{dxdz} |d_{x\uparrow}d_{z\uparrow}\rangle + L'_{dxa} |d_{x\uparrow}a_\uparrow\rangle + L'_{dxpz} |d_{x\uparrow}p_{z\uparrow}\rangle + ...,$$

$$\left| B_1^{T-1} \right\rangle = L'_{dxdz} |d_{x\downarrow}d_{z\downarrow}\rangle + L'_{dxa} |d_{x\downarrow}a_\downarrow\rangle + L'_{dxpz} |d_{x\downarrow}p_{z\downarrow}\rangle + ...$$

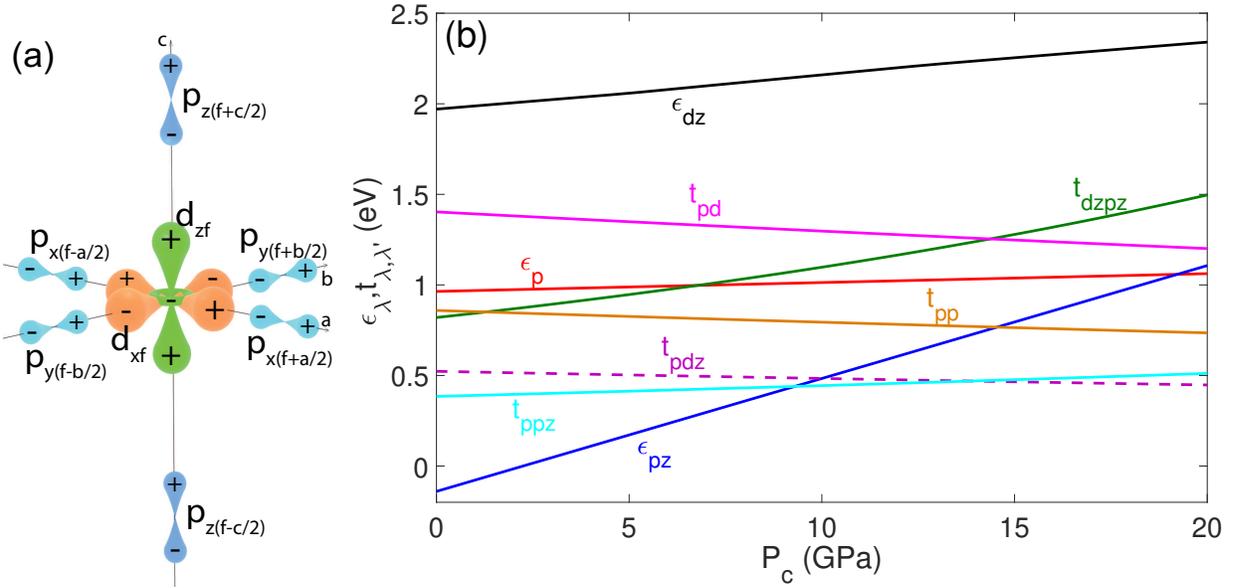

FIG. 1: (Color online) (a) Basic orbitals of the five-band p-d model in the $CuO_6$ octahedron with phase convention. (b) Dependences of on-site energies and hopping integrals on the magnitude of uniaxial pressure along the $c$ axis.

The triplet $\left|B_1^T\right\rangle$ is 1.06 eV above the ground state $\left|A_1^S\right\rangle$. The singlet component of the states that form the triplet $\left|B_1^T\right\rangle$ is slightly higher in energy than this triplet (by 0.05 eV) (Fig. 2, left panel, green line):

$$
\begin{aligned}
\left|B_2^S\right\rangle =\ & L''_{dxdz}\left(\left|d_{x\downarrow}d_{z\uparrow}\right\rangle-\left|d_{x\uparrow}d_{z\downarrow}\right\rangle\right)+L''_{dxa}\left(\left|d_{x\downarrow}a_\uparrow\right\rangle-\left|d_{x\uparrow}a_\downarrow\right\rangle\right)+L''_{dxpz}\left(\left|d_{x\downarrow}p_{z\uparrow}\right\rangle-\left|d_{x\uparrow}p_{z\downarrow}\right\rangle\right)+ \\
& L''_{bdz}\left(\left|b_\downarrow d_{z\uparrow}\right\rangle-\left|b_\uparrow d_{z\downarrow}\right\rangle\right)+L''_{ba}\left(\left|b_\downarrow a_\uparrow\right\rangle-\left|b_\uparrow a_\downarrow\right\rangle\right)+L''_{bpz}\left(\left|b_\downarrow p_{z\uparrow}\right\rangle-\left|b_\uparrow p_{z\downarrow}\right\rangle\right)
\end{aligned}
\tag{6}
$$

The splitting between the singlet $\left|B_2^S\right\rangle$ and the triplet component $\left|B_1^{T0}\right\rangle$ is due to the Hund's exchange. The singlet $\left|A_2^S\right\rangle$ (Fig. 2, left panel, blue line) which is the sum of basis states that are products of $a_{1g}$ symmetry orbitals

$$
\begin{aligned}
\left|A_2^S\right\rangle =\ & L_{dza}\left(\left|d_{z\downarrow}a_\uparrow\right\rangle-\left|d_{z\uparrow}a_\downarrow\right\rangle\right)+ \\
& L_{dzpz}\left(\left|d_{z\downarrow}p_{z\uparrow}\right\rangle-\left|d_{z\uparrow}p_{z\downarrow}\right\rangle\right)+ \\
& L_{apz}\left(\left|a_\downarrow p_{z\uparrow}\right\rangle-\left|a_\uparrow p_{z\downarrow}\right\rangle\right)+L_{dzdz}\left|d_{z\downarrow}d_{z\uparrow}\right\rangle+ \\
& L_{aa}\left|a_\downarrow a_\uparrow\right\rangle+L_{pzpz}\left|p_{z\downarrow}p_{z\uparrow}\right\rangle
\end{aligned}
\tag{7}
$$

is higher in energy than the singlet $\left|B_2^S\right\rangle$ by 1.1 eV. The singlet $\left|A_2^S\right\rangle$ is significantly lower in energy than the corresponding triplet (by 2.23 eV), the splitting of these states is proportional to the hopping integral $t_{pdz}$.

The annihilation and creation operators for basis orbitals are expressed in terms of the Hubbard X-operators $X_{\mathbf{f}}^{pq}$ which describe the quasiparticle excitations ($pq$) between the initial $\left|q\right\rangle$ and final $\left|p\right\rangle$ eigenstates of the $CuO_6$ cluster at site $\mathbf{f}$. So, for example, for the copper orbital $d_\lambda$:

$$
d_{\lambda\mathbf{f}\sigma}=\sum_{pq}\left\langle p\right|d_{\lambda\mathbf{f}\sigma}\left|q\right\rangle X_{\mathbf{f}}^{pq}=\sum_{pq}\gamma_{d\lambda\sigma}\left(pq\right)X_{\mathbf{f}}^{pq}
\tag{8}
$$

It is necessary to consider the dispersion of all possible the Fermi-type transitions ($pq$) between multiparticle eigenstates of the $CuO_6$ cluster to reproduce the complete electronic structure of the five-band p-d model in terms of quasiparticle excitations. Some quasiparticles such as the transitions $\left(b_{1\uparrow}B_1^{T-1}\right)$, $\left(b_{1\uparrow}A_2^S\right)$ have zero spectral weight due to the zero matrix elements $\gamma_{dx\uparrow}\left(b_{1\uparrow}B_1^{T-1}\right)$, $\gamma_{b\uparrow}\left(b_{1\uparrow}B_1^{T-1}\right)$, $\gamma_{dx\uparrow}\left(b_{1\uparrow}A_2^S\right)$, $\gamma_{b\uparrow}\left(b_{1\uparrow}A_2^S\right)$. To reproduce the electronic structure of quasiparticle excitations near the Fermi level, it is sufficient to take into account the following cluster eigenstates: the zero-hole state $\left|0\right\rangle$, the two single-hole states $\left|b_1\right\rangle=\left\{\left|b_{1\uparrow}\right\rangle,\left|b_{1\downarrow}\right\rangle\right\}$ and the four two-hole states $\left|L\right\rangle=\left\{\left|A_1^S\right\rangle,\left|B_1^{T0}\right\rangle,\left|B_1^{T1}\right\rangle,\left|B_2^S\right\rangle\right\}$. The basis of the multiband Hubbard model builded on these eigenstates includes the five quasiparticle excitations: $q_0\equiv\left(0b_{1\sigma}\right),q_1\equiv\left(b_{1\bar\sigma}A_1^S\right),q_2\equiv\left(b_{1\bar\sigma}B_1^{T0}\right),q_3\equiv\left(b_{1\sigma}B_1^{T1}\right),q_4\equiv\left(b_{1\bar\sigma}B_1^S\right)$ (Fig. 3). The quasiparticle $q_1$ is formed by the $d_x$- and $b$-orbitals, the quasiparticles $q_2$, $q_3$, $q_4$ are formed by the orbitals $d_z$, $a$, $p_z$. The Hamiltonian of the effective five-band Hubbard model for quasiparticle excitations in terms of Hubbard operators has the form:

$$H = \sum_{\mathbf{f}} \varepsilon_0 X_{\mathbf{f}}^{00} + \sum_{\mathbf{f}\sigma} (\varepsilon_1 - \mu) X_{\mathbf{f}}^{b_{1\sigma} b_{1\sigma}} + \sum_{\mathbf{f}L} (\varepsilon_{2L} - 2\mu) X_{\mathbf{f}}^{LL} +$$
$$\sum_{\mathbf{fg}\sigma} t_{\mathbf{fg}} (b_{1\sigma}0, 0b_{1\sigma}) X_{\mathbf{f}}^{b_{1\sigma}0} X_{\mathbf{g}}^{0b_{1\sigma}} + \sum_{\mathbf{fg}\sigma\sigma'LL'} t_{\mathbf{fg}} (Lb_{1\sigma}, b_{1\sigma'}L') X_{\mathbf{f}}^{Lb_{1\sigma}} X_{\mathbf{g}}^{b_{1\sigma'}L'} +$$
$$\sum_{\mathbf{fg}\sigma L} t_{\mathbf{fg}} (Lb_{1\sigma}, 0b_{1\sigma}) X_{\mathbf{f}}^{Lb_{1\sigma}} X_{\mathbf{g}}^{0b_{1\sigma}} + \sum_{\mathbf{fg}\sigma L} t_{\mathbf{fg}} (b_{1\sigma}0, b_{1\sigma}L) X_{\mathbf{f}}^{b_{1\sigma}0} X_{\mathbf{g}}^{b_{1\sigma}L} \tag{9}$$

Here the hopping integrals for the quasiparticle excitations $(pq)$ and $(mn)$ are expressed in terms of the hopping integrals in the Hamiltonian (II) as follows:

$$t_{\mathbf{fg}} (pq, mn) = (-1) \times \tag{10}$$
$$2t_{pd}\mu_{\mathbf{fg}} \left( \gamma_{dx\sigma}^* (pq) \gamma_{b\sigma} (mn) + \gamma_{b\sigma}^* (pq) \gamma_{dx\sigma} (mn) \right) +$$
$$2t_{pdz}\xi_{\mathbf{fg}} \left( \gamma_{dz\sigma}^* (pq) \gamma_{b\sigma} (mn) + \gamma_{b\sigma}^* (pq) \gamma_{dz\sigma} (mn) \right) -$$
$$2t_{pdz}\lambda_{\mathbf{fg}} \left( \gamma_{dz\sigma}^* (pq) \gamma_{a\sigma} (mn) + \gamma_{a\sigma}^* (pq) \gamma_{dz\sigma} (mn) \right) +$$
$$2t_{pp}\nu_{\mathbf{fg}} \left( -\gamma_{b\sigma}^* (pq) \gamma_{b\sigma} (mn) + \gamma_{a\sigma}^* (pq) \gamma_{a\sigma} (mn) \right) -$$
$$2t_{pp}\chi_{\mathbf{fg}} \left( \gamma_{b\sigma}^* (pq) \gamma_{a\sigma} (mn) + \gamma_{a\sigma}^* (pq) \gamma_{b\sigma} (mn) \right) +$$
$$2t_{ppz}\xi_{\mathbf{fg}} \left( \gamma_{b\sigma}^* (pq) \gamma_{pz\sigma} (mn) + \gamma_{pz\sigma}^* (pq) \gamma_{b\sigma} (mn) \right) -$$
$$2t_{ppz}\lambda_{\mathbf{fg}} \left( \gamma_{a\sigma}^* (pq) \gamma_{pz\sigma} (mn) + \gamma_{pz\sigma}^* (pq) \gamma_{a\sigma} (mn) \right)$$

where the structural factors $\nu_{\mathbf{fg}}$, $\lambda_{\mathbf{fg}}$, $\xi_{\mathbf{fg}}$, $\chi_{\mathbf{fg}}$ are determined in $k$-space in [56, 61].

The method of equations of motion for Green's functions of quasiparticle excitations is used to obtain the electronic structure. The Green's functions built on creation and annihilation Fermi operators is expressed in terms of Green's functions built on Hubbard operators by the following expression (using a example of the copper $d$-orbitals):

$$\left\langle \left\langle d_{\lambda\mathbf{f}\sigma}(t) \mid d_{\lambda'\mathbf{f}'\sigma}^{\dagger}(0) \right\rangle \right\rangle = \sum_{pqmn} \gamma_{d\lambda\sigma}^* (pq) \gamma_{d\lambda'\sigma} (mn) \left\langle \left\langle X_{\mathbf{f}}^{pq}(t) \mid X_{\mathbf{f}'}^{mn}(0) \right\rangle \right\rangle \tag{11}$$

The equations of motion for the components of the matrix form of the Green's function $\hat{D}(\mathbf{f}, \mathbf{f}'; t)$ have the general form:

$$\omega D_{(pq)(mn)}(\mathbf{f}, \mathbf{f}'; t) = \delta_{pn}\delta_{qm} F(pq) + \Omega(pq) D_{(pq)(mn)}(\mathbf{f}, \mathbf{f}'; t) + Z_{(pq)(mn)}(\mathbf{f}, \mathbf{f}'; t) \tag{12}$$

where the term $Z_{(pq)(mn)}(\mathbf{f}, \mathbf{f}'; t) = \langle\langle Z_{\mathbf{f}}^{pq}(t) \mid X_{\mathbf{f}'}^{mn}(0)\rangle\rangle$ includes high-order Green's functions. $\Omega(pq) = \varepsilon_p - \varepsilon_q - \mu$ is the energy of the quasiparticle $(pq)$. The quantities $F(pq) = \langle X^{pp}\rangle + \langle X^{qq}\rangle$ are the filling factors with the filling numbers $\langle X^{pp}\rangle$ of the state $p$. Decoupling of the equations of motion is performed in the generalized mean-field approximation taking into account the spin-spin $C_{\mathbf{fg}} = \left\langle X_{\mathbf{f}}^{b_\sigma b_\sigma} X_{\mathbf{g}}^{b_\sigma b_\sigma} \right\rangle$ and the kinematic $K_{\mathbf{fg}} = \left\langle X_{\mathbf{f}}^{b_{1\sigma}0} X_{\mathbf{g}}^{0b_{1\sigma}} \right\rangle$, $\left\langle X_{\mathbf{f}}^{b_{1\sigma}0} X_{\mathbf{g}}^{b_{1\sigma}L} \right\rangle$, $\left\langle X_{\mathbf{f}}^{Lb_{1\sigma}} X_{\mathbf{g}}^{0b_{1\sigma}} \right\rangle$, $\left\langle X_{\mathbf{f}}^{Lb_{1\sigma}} X_{\mathbf{g}}^{b_{1\sigma}L} \right\rangle$ correlation functions, where $L = A_1^S, B_1^{T0}, B_1^{T1}, B_2^S$, using a projection technique of the Mori type [62]. In this decoupling method, the operators $Z_{\mathbf{f}}^{pq}$ are represented as a sum of reducible and irreducible parts [62]. The reducible part can be linearized with respect to X-operators of the considered model of

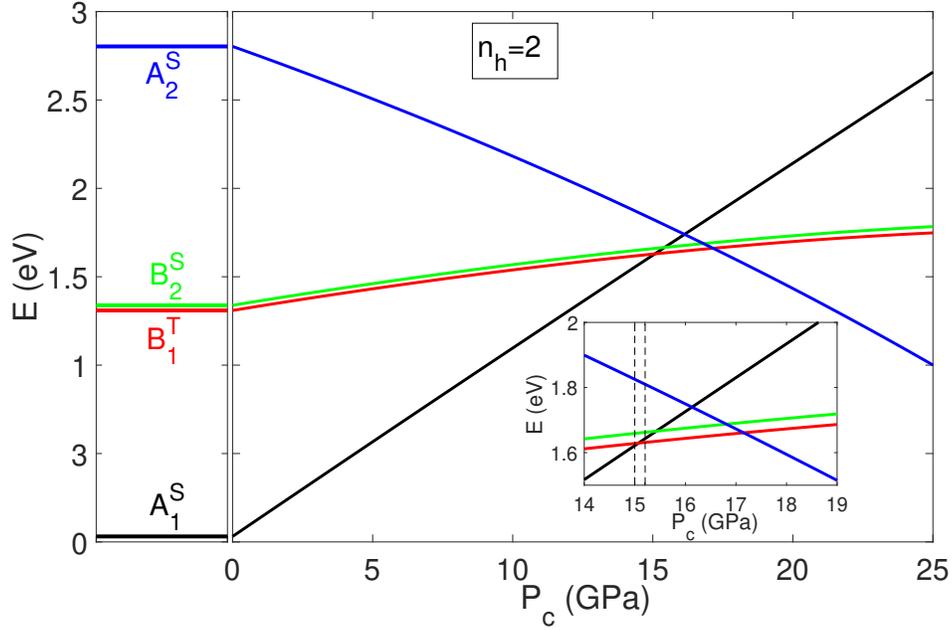

FIG. 2: (Color online) (Left panel) The levels of the six (the level $|B_1^T\rangle$ is threefold degenerate) lowest-energy two-hole eigenstates of the $CuO_6$ cluster. (Right panel) The dependences of the energies of two-particle states on the magnitude of uniaxial pressure. The inset shows the same energies of two-hole states as in the main figure but in the crossover region. The dotted lines in the inset indicate the pressure values before and after the crossover at which the band structure will be calculated.

quasiparticle excitations:

$$Z_{\mathbf{f}}^{pq} = \sum_{pq} T_{\mathbf{f}}^{(pq)(q'p')} X_{\mathbf{f}}^{(q'p')} + Z_{\mathbf{f}}^{(pq)(irr)} \qquad (13)$$

where $T_{\mathbf{j}}^{(pq)(q'p')} = \frac{\left\langle \left\{ Z_{\mathbf{j}}^{pq}, X_{\mathbf{f}}^{p'q'} \right\} \right\rangle}{\left\langle \left\{ X_{\mathbf{f}}^{p'q'}, X_{\mathbf{f}}^{q'p'} \right\} \right\rangle}$ are the linearization coefficients that define the self-energy operator. The kinematic correlators are calculated self-consistently together with the Fermi level and the occupation numbers of local cluster eigenstates $\langle X_{\mathbf{f}}^{00}\rangle$, $\langle X_{\mathbf{f}}^{b_{1\sigma}b_{1\sigma}}\rangle$, $\langle X_{\mathbf{f}}^{b_{1\sigma}b_{1\sigma}}\rangle$, $\langle X_{\mathbf{f}}^{LL}\rangle$, where $L = A_1^S$, $B_1^{T0}$, $B_1^{T1}$, $B_2^S$ [63]. The values of the spin-spin correlation functions were calculated within the framework of the t-J* model in [64, 65] taking into account only ground two-hole singlet state $|A_1^S\rangle$. The spin-spin correlator values may change if the triplet states are taken into account. It was shown in [66] that the overwhelming contribution to the effective exchange interaction in the $CuO_2$ plane comes from hoppings of the excitations involving the singlet state $|A_1^S\rangle$; this contribution is of the antiferromagnetic type. The contribution to the superexchange interaction that is caused by the hoppings of the quasiparticles involving the triplet states is only 3%. Accordingly, the influence of the exchange interaction caused by the triplet states on spin correlations will be small. These conclusions are valid even if the triplet state becomes the ground state [66].

We obtain the Dyson equation in $k$-space neglecting the irreducible operator $Z_{\mathbf{f}}^{(pq)(irr)}$:

$$\hat{D}(\mathbf{k}; \omega) = \left( \omega - \hat{\Omega} - \hat{F}\hat{T}_{\mathbf{k}} \right)^{-1} \qquad (14)$$

where

$$\hat{\Omega} = diag\left( \Omega(0b_{1\sigma}), \Omega\left(b_{1\bar{\sigma}}A_1^S\right), \Omega\left(b_{1\bar{\sigma}}B_1^{T0}\right), \Omega\left(b_{1\sigma}B_1^{T1}\right), \Omega\left(b_{1\bar{\sigma}}B_1^S\right) \right) \qquad (15)$$

is the matrix of quasiparticle energies,

$$\hat{F} = diag\left(F\left(0b_{1\sigma}\right), F\left(b_{1\sigma}A_1^S\right), F\left(b_{1\sigma}B_1^{T0}\right), F\left(b_{1\sigma}B_1^{T1}\right), F\left(b_{1\sigma}B_1^S\right)\right) \quad (16)$$

is the matrix of filling factors, and $\hat{T}_{\mathbf{k}}$ is the matrix form of the self-energy operator with the elements $T_{\mathbf{k}}^{(pq)(q'p')}$ for five quasiparticles $(pq)$.

## IV. THE BAND STRUCTURE OF QUASIPARTICLE EXCITATIONS WITHIN THE FRAMEWORK OF THE FIVE-BAND EFFECTIVE HUBBARD MODEL WITHOUT PRESSURE

The quasiparticle $q_0$ (Figs. 3a-d) forms the conduction band that is in the range from 2 to 4 eV (Fig. 3d). The quasiparticles $q_1$, $q_2$, $q_3$, and $q_4$ participate in the formation of the upper part of the valence band (Fig. 3d). The Fermi level crosses the band that is mainly formed by the quasiparticle $q_1$ at $P_c = 0$ GPa, this band $q_1$ lies between $-2.5$ and $0.5$ eV (Fig. 3d). All quasiparticles contribute to each of the bands, but the predominant contribution of a certain quasiparticle can be distinguished for some bands. The band is designated by the quasiparticle name in the sense of the predominant contribution of this quasiparticle to this band. The maximum of the valence band is at point $(\pi/2, \pi/2)$. The band dispersion and the Fermi contour have different forms for different doping levels. The energy of the state in the local minimum at the $k$-point $(\pi, \pi)$ and its vicinity increases with increasing doping. Another characteristic change with increasing doping is a more pronounced maximum in the $k$-direction $(\pi, \pi)$ - $(\pi, 0)$. The Fermi level falls deeper into the valence band as hole doping increases. The transformation of the Fermi contour under doping is determined by two factors: a dispersion renormalization and a change in the Fermi level position. The shape of the Fermi contour obtained here within the framework of the fiveband effective Hubbard model repeats the shape obtained within the framework of the Hubbard model with two quasiparticles. The Fermi contour consists of four hole pockets around the points $(\pi/2, \pi/2)$, $(3\pi/2, \pi/2)$, $(\pi/2, 3\pi/2)$, $(3\pi/2, 3\pi/2)$ at $x = 0.1$ (Fig. 8a). The Fermi contour at $x = 0.15$ represents four almost connected hole pockets (Fig. 9a). The Fermi contour consists at $x = 0.25$ of a large outer contour and a smaller inner contour around the point $(\pi, \pi)$ which are formed by the merging of the four hole pockets (Fig. 10a). The bands of quasiparticles $q_2$, $q_3$, $q_4$ have a weak effect on the Fermi contour at $P_c = 0$ GPa. The quasiparticle bands $q_2$-$q_3$-$q_4$ are between $-3$ and $-1.5$ eV and cross the lower part of the $q_1$ band (Figs. 3d, 5a, 6a, 7a). The position and dispersion of the bands $q_2$-$q_3$-$q_4$ for different hole concentrations change slightly but no fundamental reconstruction occurs. Only the states of one of these three bands have a significant spectral weight at $P_c = 0$ GPa, as can be seen from Figs. 5a, 6a, 7a. The predominant contribution to this band is made by the quasiparticle $q_3$; the hybridization with the quasiparticles $q_2$ and $q_4$ changes its dispersion although the general shape is preserved.

Each quasiparticle contains contributions from a certain orbitals. As a result, each band contains the contributions of the five orbitals $\lambda = d_x, b, p_z, d_z, a$:

$$A_\lambda\left(\mathbf{k}, \omega + i\delta\right) = -\frac{1}{\pi} \sum_{pqmn} \gamma_\lambda\left(pq\right) \gamma_\lambda^*\left(nm\right) \mathrm{Im}\left(\langle\langle X_{\mathbf{k}}^{pq} \mid X_{\mathbf{k}}^{mn}\rangle\rangle_{\omega+i\delta}\right). \quad (17)$$

The magnitudes of $A_\lambda\left(\mathbf{k}, \omega + i\delta\right)$ depend on a wave vector and pressure (Fig. 4). At zero pressure, the band $q_1$ is formed predominantly by the $d_x$- and $b$-orbitals (Fig. 4a), the contribution of the orbitals $d_z$, $a$, $p_z$ is much smaller although the contribution of the $p_z$-orbital is noticeable for states with wave vectors in the $(\pi, 0)$ - $(0, 0)$ direction (Fig. 4c). The $q_2$-$q_3$-$q_4$ bands are formed mainly by the $p_z$-orbital of the apical oxygen, the contributions of the copper $d_z$-orbital and molecular $a$-orbital of the planar oxygens are much smaller.

## V. THE INFLUENCE OF UNIAXIAL PRESSURE ALONG THE $c$ AXIS ON THE ELECTRONIC STRUCTURE OF LOCAL STATES AND QUASIPARTICLE EXCITATIONS

### A. The influence of pressure on the structure of local many-particle states

The ground single-hole and two-hole states are formed by the $b_{1g}$ symmetry orbitals. Uniaxial compression along the $c$ axis increases the hybridization of $p_z$-orbitals with planar orbitals, reduces tetrahedral distortion and splitting of the energy levels of $d_x$- and $d_z$-orbitals. This leads to a decrease in the energy difference of the ground state and the excited states including $a_{1g}$ symmetry orbitals. If we consider the large scale of changes in uniaxial pressure the crossover between the two-hole ground state $\left|A_1^S\right\rangle$ (Fig. 2 black line) and the singlet of $a_{1g}$ orbitals $\left|A_2^S\right\rangle$ (the fifth excited state, Fig. 2 blue line) is clearly visible, while the triplet state $\left|B_1^T\right\rangle$ together with the singlet state $\left|B_1^S\right\rangle$ remain between singlet $\left|A_1^S\right\rangle$ and singlet $\left|A_2^S\right\rangle$. However, it can be seen from Fig. 2 that initially a crossover occurs between $\left|A_1^S\right\rangle$ and the triplet $\left|B_1^T\right\rangle$ (the first excited state, Fig. 2, red line) at a value $P_c \approx 15.1$ GPa. The triplet remains the ground state in a small range of pressure values. The ground state becomes a singlet $\left|A_2^S\right\rangle$ when $P_c$ increasing above 17.2 GPa. Further we will consider the crossover between the singlet $\left|A_1^S\right\rangle$ and the triplet $\left|B_1^T\right\rangle$.

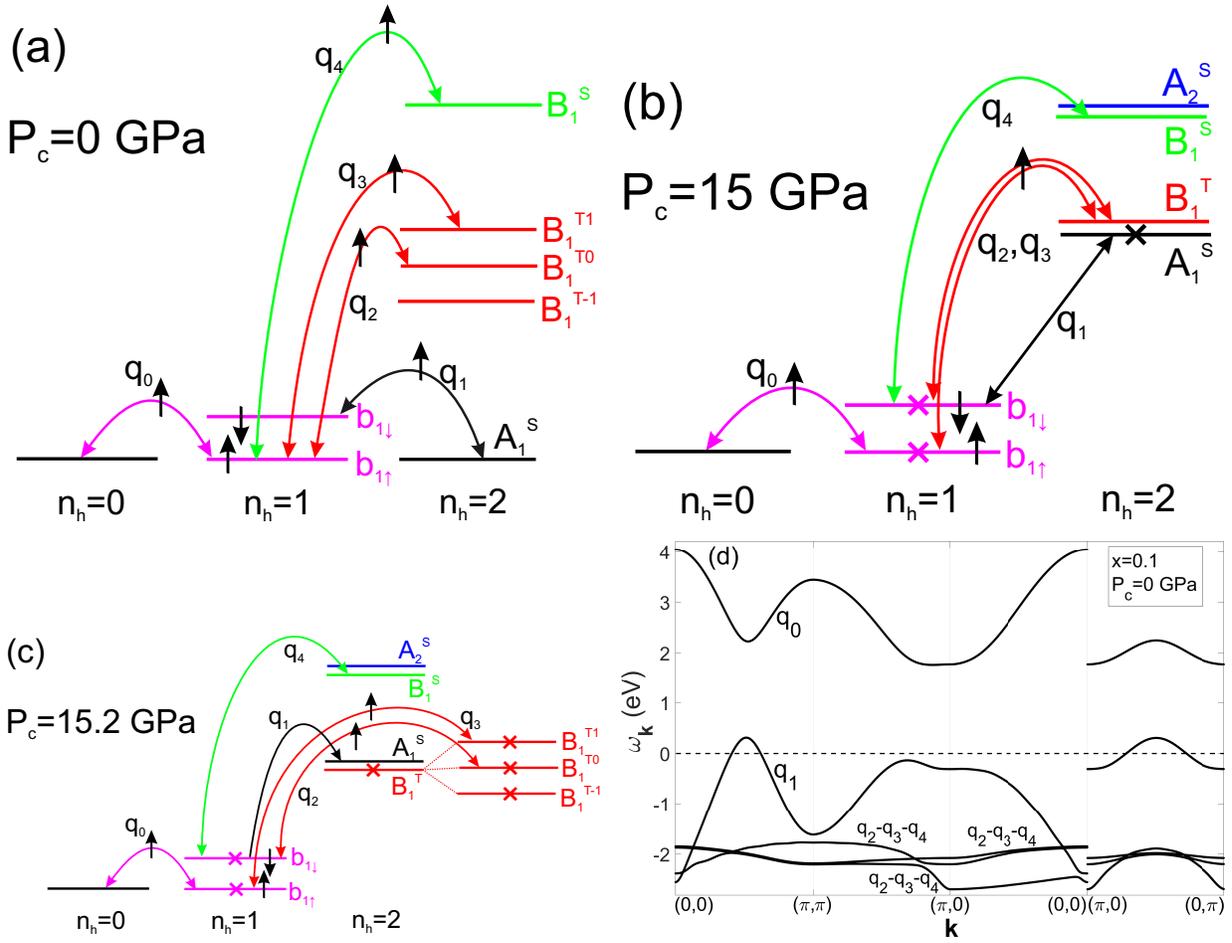

FIG. 3: (Color online) The general structure of local quasiparticle excitations between many-particle eigenstates of the CuO$_6$ cluster that form the top of the valence band at values of compression along the $c$ axis less (a,b) and greater (c) than the pressure $P_c = 15.2$ GPa at which crossover occurs between local copper-oxygen singlet and triplet states. (d) The band structure of quasiparticle excitations without spectral weight at $x = 0.1$.

The crossover of the single-hole states $\left|b_1\right\rangle$ and $\left|a_1\right\rangle$ with $P_c$ variation occurs approximately at the value $P_c = 21$ GPa. Therefore, the doublet $\left|b_1\right\rangle$ remains the ground state in the single-hole sector in the range of values $P_c =$ 15.1÷21 GPa. This means that all effects in this pressure range are due to the crossover of two-hole states.

## B. The general effect of uniaxial pressure on the band structure and the Fermi contour of quasiparticle excitations

Growth of the hybridization of the apical oxygen $p_z$-orbitals with planar oxygen orbitals under the $c$ axis compression leads to an increase in the energy of electron $a_{1g}$ symmetry orbitals. This pushes bands of the excitations $q_2$, $q_3$, $q_4$ to the top of the electron valence band. At $P_c = 10$ GPa and hole concentrations $x = 0.1$ and $x = 0.15$, the maximum of the band $q_2$-$q_3$-$q_4$ at the point $(\pi, \pi)$ rises above the minimum of the band $q_1$ at this point (Figs. 5b, 6b). The hybridization of the quasiparticles $q_2$, $q_3$, $q_4$ with the quasiparticle $q_1$ increases as the energy of the $q_2$-$q_3$-$q_4$ bands increases. Hybridization of the quasiparticle $q_1$ with the quasiparticles $q_2$, $q_3$, $q_4$ results in the fact that states of the band $q_1$ in the vicinity

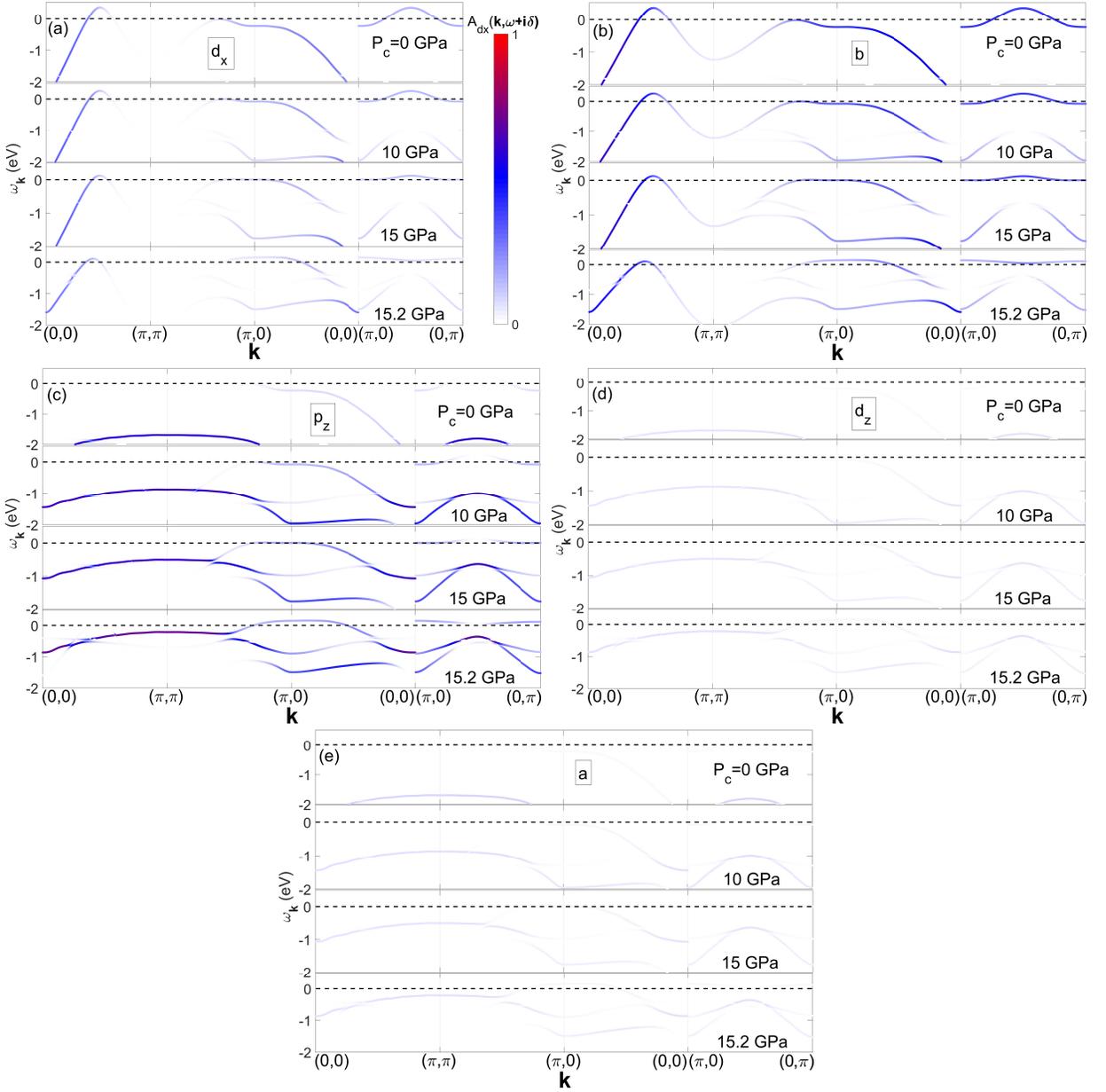

FIG. 4: (Color online) The contributions $A_\lambda (\mathbf{k}, \omega + i\delta)$ (Eq. 17) of the five orbitals $\lambda = d_x, b, p_z, d_z, a$ to the spectral weight of the valence band states at $x = 0.15$ and different magnitudes of uniaxial compression

of the local maximum near the point $(\pi, 0)$ become a part of the upper of the $q_2$-$q_3$-$q_4$ bands, and vice versa, the band $q_1$ acquires states deep in the valence band which previously belonged to the $q_2$-$q_3$-$q_4$ bands. The latter effect is more clearly visible at the pressure $P_c = 15$ GPa in Figs. 5c, 6c. There is also a redistribution of the spectral weight between the split bands $q_2$-$q_3$-$q_4$: the states of the band with a predominant contribution from the quasiparticle $q_4$, which do not have spectral weight at $P_c = 0$ GPa, acquire weight with increasing pressure. This effect is seen from how states with energies from $-1.3$ to $-1$ eV are highlighted between the band $q_1$ and the high-intensity band $q_2$-$q_3$-$q_4$ in the directions $(\pi, \pi) - (\pi, 0)$ and $(\pi, 0) - (0, 0)$ at $P_c = 10$ GPa at $x = 0.1$, 0.15 (Figs. 5b, 6b). These states with increasing intensity stand out in a separate band due to strong splitting with other high-intensity bands at $P_c = 15$ and $P_c = 15.2$ GPa (Figs. 5c,d, 6c,d, 7d).

An increase in the contribution of quasiparticles $q_2$, $q_3$, $q_4$ to the region of states on the top of the valence band in the vicinity of the point $(\pi, 0)$ with increasing pressure occurs simultaneously with an increase in the contribution of the $p_z$-orbital to these and other states which is clearly noticeable in the $(\pi, 0)$-$(0, 0)$ direction (Fig. 4c). In this case, the contributions of $d_x$- and $b$-orbital in these states decrease (Figs. 4a,b). The energy of quasiparticle excitations with wave vectors in the vicinity of $(\pi, 0)$ increases with increasing compression (Figs. 5b, 6b) so that a broad maximum is formed there (Figs. 5c, 6c). This effect is clearly seen at the compression $P_c = 15$ GPa for the hole concentrations $x = 0.1$ and 0.15. The wide maximum in the vicinity of $(\pi, 0)$ is also formed for the hole concentration $x = 0.25$ but at a pressure greater than that required for the singlet-triplet crossover; this is no longer a local maximum but a global one (Fig. 7d). The energy of states of the band $q_1$ with wave vectors near the point $(\pi/2, \pi/2)$ decreases with increasing compression.

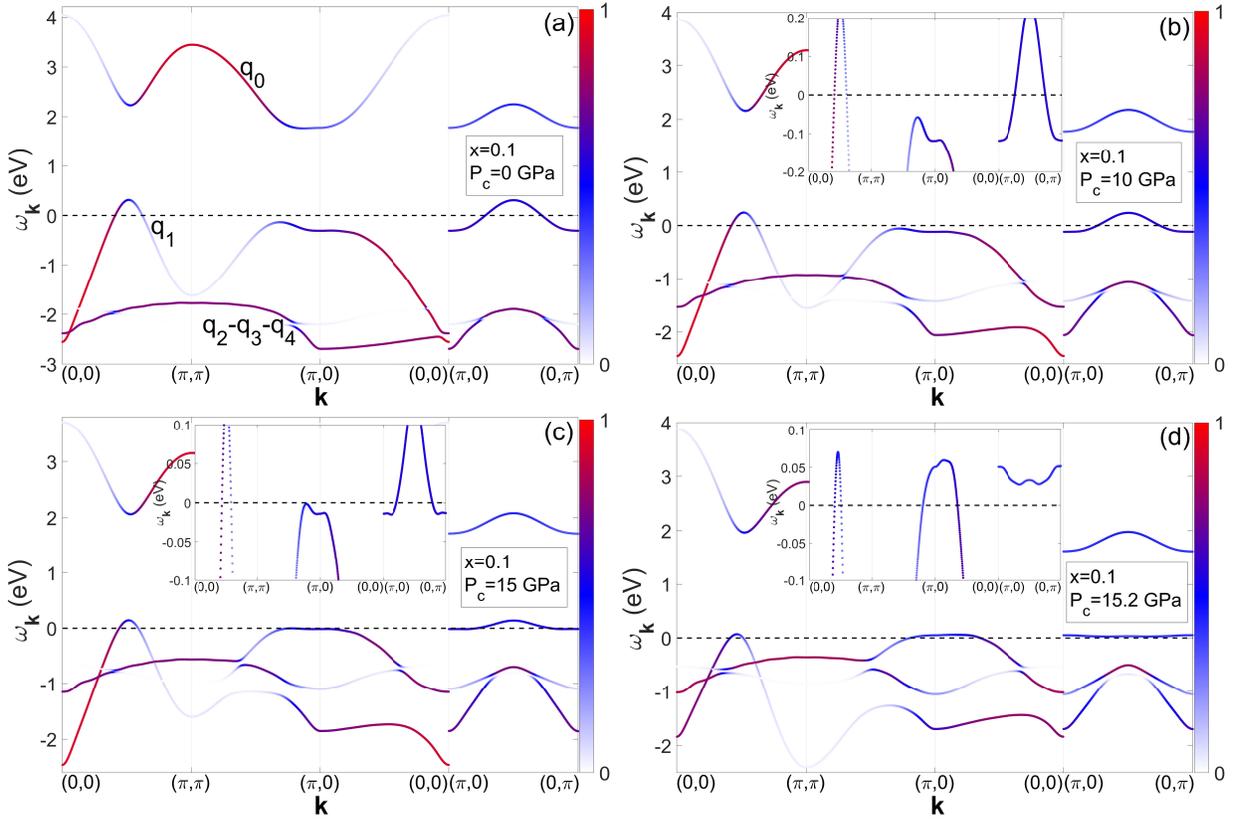

FIG. 5: (Color online) Band structure of the five quasiparticle excitations for the doped hole concentration $x = 0.1$ at different values of uniaxial compression $P_c$.

For both underdoped and overdoped compositions, an increase in uniaxial pressure up to crossover values does not lead to significant transformations of the Fermi contour. The Fermi contour at $x = 0.1$ has the form of the four hole pockets up to the crossover at $P_c = 15.1$ GPa (Figs. 8a,b), only the shape of these pockets changes. The Fermi contour at $x = 0.25$ has the form of the outer and inner contours up to the crossover value (Figs. 10a,b). Changes in the Fermi surface topology can occur at pressure values significantly lower than those required for the singlet-triplet crossover only under a certain condition. This condition is the proximity of the Fermi level to the band region where there are such dispersion features as local maxima and minima. In $La_{2-x}Sr_xCuO_4$, this specific condition is satisfied at $x = 0.15$ when the

Fermi level almost touches the local maximum in the $(\pi, \pi)$-$(\pi, 0)$ direction (Fig. 6b). In this case, even minor changes in pressure causing a minor renormalization of the dispersion lead to a radical change in the Fermi contour.

The region of the broad maximum in the vicinity $(\pi, 0)$ at a large energy scale resembles a flat band (Figs. 5b, 6b,c) but a certain dispersion shape is clearly seen at a small energy scale (the insets of Figs. 5b,c, 6b,c). The band $q_1$ structure changes slightly on a large energy scale with varying pressure from 10 to 15 GPa but the band $q_1$ reconstruction is very significant on a small scale near the Fermi level, as can be seen in the insets of Figs. 5b,c, 6b,c for the hole concentration $x = 0.15$. It is the reconstruction of the weak, but clearly expressed on a small energy scale, dispersion of the band $q_1$ near the Fermi level that leads to changes in the Fermi contour topology. There is one local maximum $M1$ of the $q_1$ band in the direction $(\pi, \pi) - (\pi, 0)$ and the barely pronounced maximum $M2$ in the direction $(\pi, 0) - (0, 0)$ (inset of Fig. 6b) at $P_c = 10$ GPa. The band $q_1$ reconstruction under compression leads to the fact that Fermi level falls on the van Hove singularity in the density of states caused by the local maximum $M1$. In this case, the Fermi contour topology changes: while the Fermi contour at $P_c = 0$ GPa represents individual hole pockets (Fig. 9a), the Fermi contour at $P_c = 10$ GPa represents two large contours, outer and inner, as a result of the merge of these pockets (Fig. 9b).

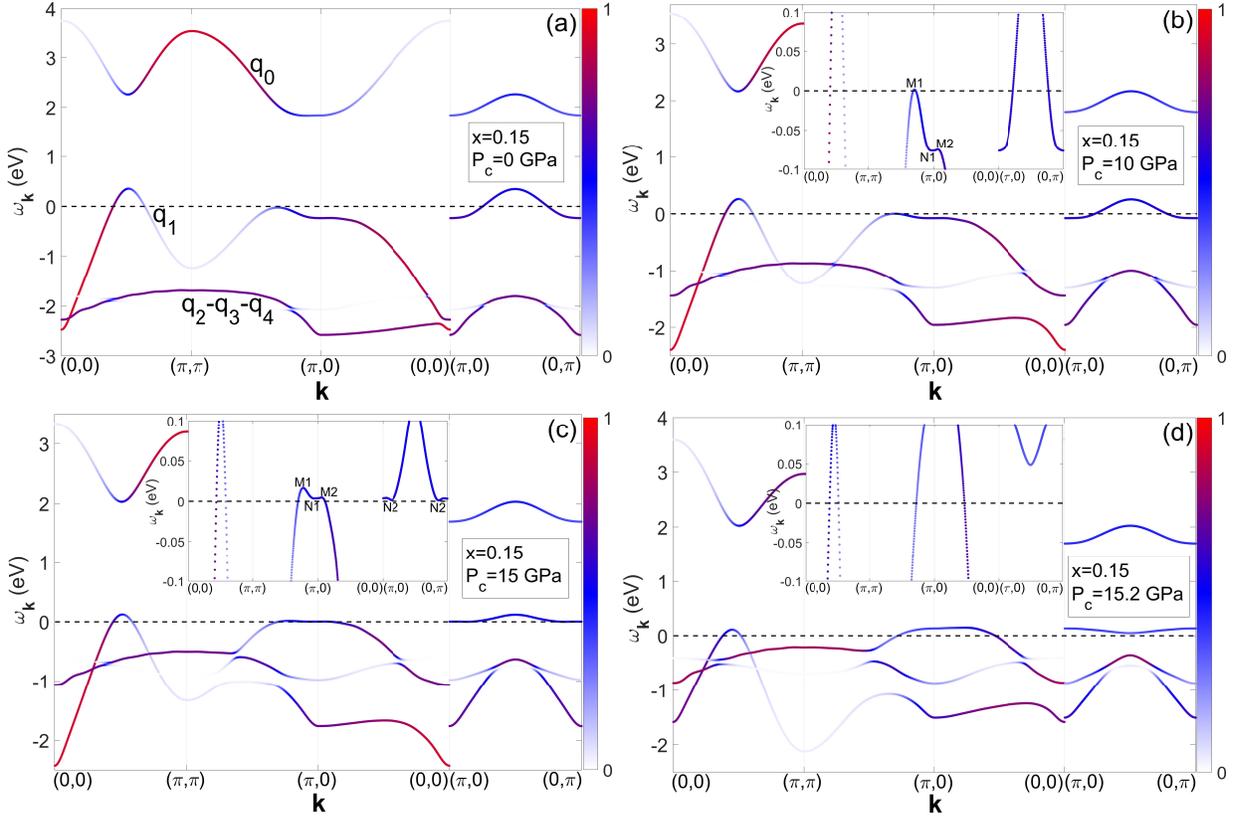

FIG. 6: (Color online) Band structure of the five quasiparticle excitations for the doped hole concentrations $x = 0.15$ at different values of uniaxial compression $P_c$.

The large outer Fermi contour increase with further increase in compression, its angles approach the edges of the Brillouin zone at the points $(\pi, 0)$, $(0, \pi)$, $(\pi, 2\pi), (2\pi, \pi)$ (Fig. 9c). The corners of each of the outer contours elongated in the direction of the points $(\pi, 0)$, $(0, \pi)$, $(\pi, 2\pi), (2\pi, \pi)$ have a very unusual fork shape with two teeth closed in an arc (Fig. 9c). Hybridization between the quasiparticle $q_1$ and the quasiparticles $q_2$, $q_3$, $q_4$ leads to a flattening of the region of states near the local maximum in the vicinity of the point $(\pi, 0)$ so that the maxima $M1$ and $M2$ are aligned (inset of Fig. 6c). The Fermi level reaches the local maximum $M2$ at $P_c$ slightly larger than 14.9 GPa and a second quantum phase transition occurs: each of the fork teeth of the large outer Fermi contour closes at the boundary of the Brillouin zone, very small pockets around the points $(\pi, 0)$, $(0, \pi)$, $(\pi, 2\pi), (2\pi, \pi)$ and four large pockets around the points $(0, 0)$, $(0, 2\pi)$, $(2\pi, 0)$, $(2\pi, 2\pi)$ are formed after this closure. It is seen from the band structures at $P_c \geq 14.9$ GPa (Fig. 6d) that the states at the top of the valence band in the region of the point $(\pi, 0)$ which belonged to the band $q_1$ at lower pressures now belong to the band of the quasiparticles $q_2$, $q_3$, $q_4$. Hybridization between the quasiparticle $q_1$ and the quasiparticles $q_2$, $q_3$, $q_4$ is strong for states with wave vectors in the vicinity of the point $(\pi, 0)$, and it is impossible to define the predominant contribution of one or another quasiparticle excitation to these states.

The third quantum phase transition occurs at slightly higher pressures when the Fermi level reaches the local minimum $N1$ (Fig. 6c): the small pockets disappear leaving only the four pockets around the points $(0, 0)$, $(0, 2\pi)$, $(2\pi, 0)$, $(2\pi, 2\pi)$ and the inner pocket around the point $(\pi, \pi)$ (Fig. 9d). This is what the Fermi contour looks like before the crossover at $P_c = 15.1$ GPa.

Thus, an increase in uniaxial pressure leads to several main trends in the transformation of the band structure. The energy of excitations in the bands $q_2$-$q_3$-$q_4$ increases throughout the Brillouin zone, this is especially clearly seen in the states in the vicinity of the point $(\pi, \pi)$. An increase in the hybridization between the quasiparticle $q_1$ and the quasiparticles $q_2$, $q_3$, $q_4$ changes the dispersion at the top of the valence band: the energy of the states in the vicinity of points $(\pi, 0)$, $(0, \pi)$, $(\pi, 2\pi)$, $(2\pi, \pi)$ increases, and the energy of the states in the vicinity of points $(\pi/2, \pi/2)$, $(3\pi/2, \pi/2)$, $(\pi/2, 3\pi/2)$, $(3\pi/2, 3\pi/2)$ decreases. In addition, the splitting between the quasiparticle bands deep in the valence band increases and the spectral weight is redistributed between them.

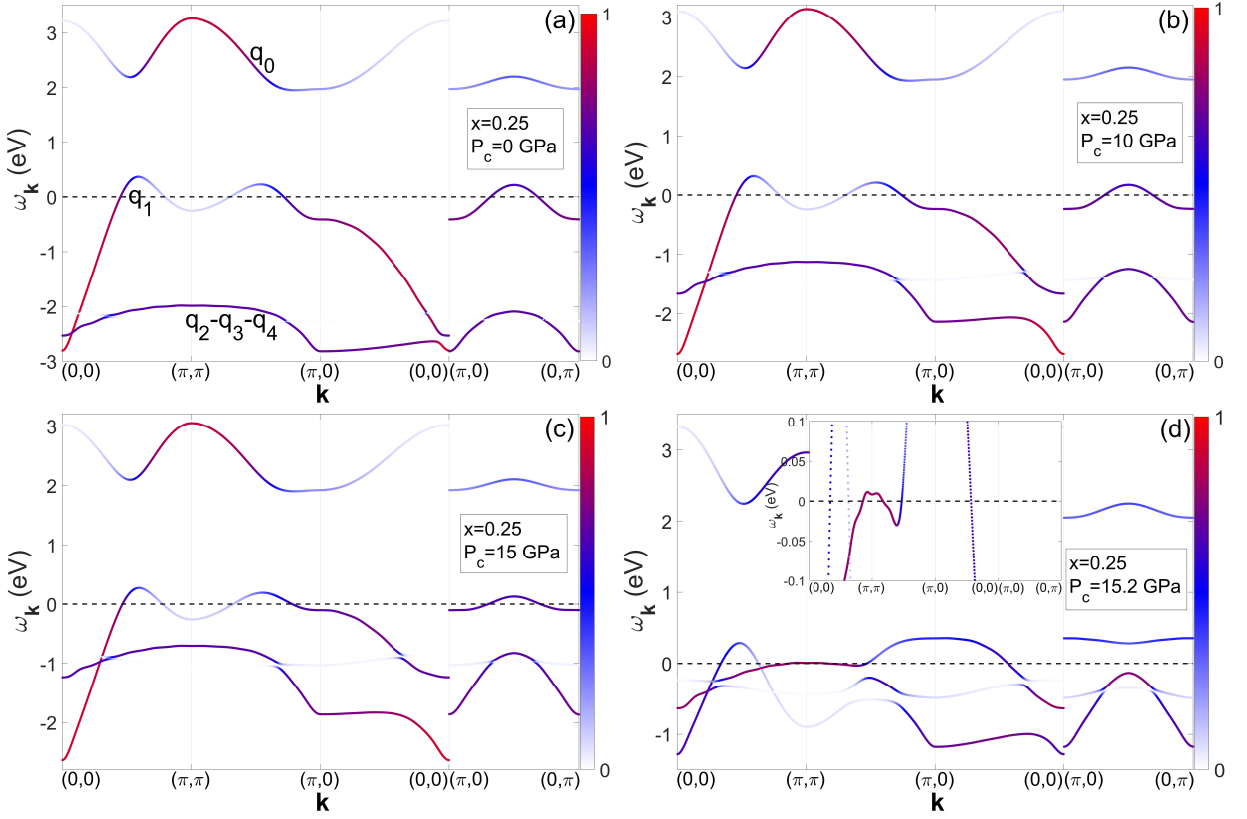

FIG. 7: (Color online) Band structure of the five quasiparticle excitations for the doped hole concentrations $x = 0.25$ at different values of uniaxial compression $P_c$.

## C. The effect of the singlet-triplet crossover on the electronic structure of quasiparticle excitations

When the Emery-Reiter triplet $\left|B_1^T\right\rangle$ becomes the ground state in the two-hole sector of Hilbert space as a result of crossover, its filling changes from 0 to $x + n_0$, where $n_0$ is the filling number of the vacuum state of holes, the filling of the singlet $\left|A_1^S\right\rangle$ becomes zero (Fig. 3c). The filling factor of the quasiparticles $q_2$ and $q_3$ increases abruptly and the intensity of the quasiparticle $q_1$ sharply decreases. The transformation processes in the band structure that developed gradually at lower pressures are realized abruptly with increasing pressure from $P_c = 15$ GPa to $P_c = 15.2$ GPa. There are few main sharp changes in the band structure that occur as a result of the singlet-triplet crossover for all three hole concentrations $x = 0.1$, $0.15$, $0.25$. Firstly, the states in the vicinity of the point $(\pi, 0)$ (Figs. 5d, 6d, 7d) reach the top of the valence band at $P_c = 15.2$ GPa (after crossover), while the absolute maximum of the valence band at $P_c = 15$ GPa (before crossover) is at the $k$-point $(\pi/2, \pi/2)$ (Figs. 5c, 6c, 7c). Accordingly, the character of the first removal state changes: the top of the valence band is formed by the $d_{x^-}$, $b$-orbitals before the crossover (Figs. 4a,b) and by the $p_z$-orbital after crossover (Fig. 10c). Secondly, the dispersion at the top of the valence band in the direction $(\pi, 0) - (0, \pi)$ just above the Fermi level is turned inside out (as can be seen from the comparison of Figs. 5c, 6c, 7c and Figs. 5d, 6d, 7d), the state at the $k$-point $(\pi/2, \pi/2)$ turns from a maximum to a minimum in the direction $(\pi, 0) - (0, \pi)$. Thirdly, the energy of the states of the band $q_2$-$q_3$-$q_4$ with wave vectors in the vicinity of the point $(\pi, \pi)$ increases significantly (Figs. 5d, 6d, 7d). Fourthly, the Fermi contour topology changes. The Fermi contour takes the form of the four pockets around the points $(0, 0)$, $(0, 2\pi)$, $(2\pi, 0)$, $(2\pi, 2\pi)$ and one rectangular pocket around $(\pi, \pi)$ (Figs. 8c, 9e, 10c) as a result of the crossover for all three concentrations. The contour of a complex shape with high-intensity states around $(\pi, \pi)$ is additionally formed at the doping $x = 0.25$ (Fig. 10c). The latter effect is due to the fact that the Fermi level for a high hole concentration is lowered deeper into the valence band than for lower hole concentrations. The Fermi level crosses the band $q_2$-$q_3$-$q_4$ in which the energy of states in the vicinity of the point $(\pi, \pi)$ increases under compression.

The transformations of the Fermi contour topology with changing pressure near the crossover at $x = 0.15$ do not occur so abruptly as for the concentrations $x = 0.1$ and $0.25$ since many of these transformations occur long before the crossover. At the concentration $x = 0.15$, even a small reconstruction of the dispersion at a compression less than what is necessary for the crossover is enough for the Fermi level to cross the regions of the band $q_2$-$q_3$-$q_4$ in the vicinity of the point $(\pi, 0)$. The gradual shift of these states relative to the Fermi level results in the fact the Fermi contour take the form of four pockets around $(0, 0)$, $(0, 2\pi)$, $(2\pi, 0)$, $(2\pi, 2\pi)$ and one square-shaped contour through a series of transformations (Fig. 9d). The curvature of the four pockets around $(0, 0)$, $(0, 2\pi)$, $(2\pi, 0)$, $(2\pi, 2\pi)$ changes as a result of the crossover (Fig. 9e). The corners of the inner square-shaped pocket around $(\pi, \pi)$ become smoother when passing the crossover, the shape of this pocket becomes more rounded (Figs. 9d,e). Thus, at $x = 0.15$, the singlet-triplet crossover does not lead to the appearance or disappearance of Fermi contour elements but the shape of existing elements changes.

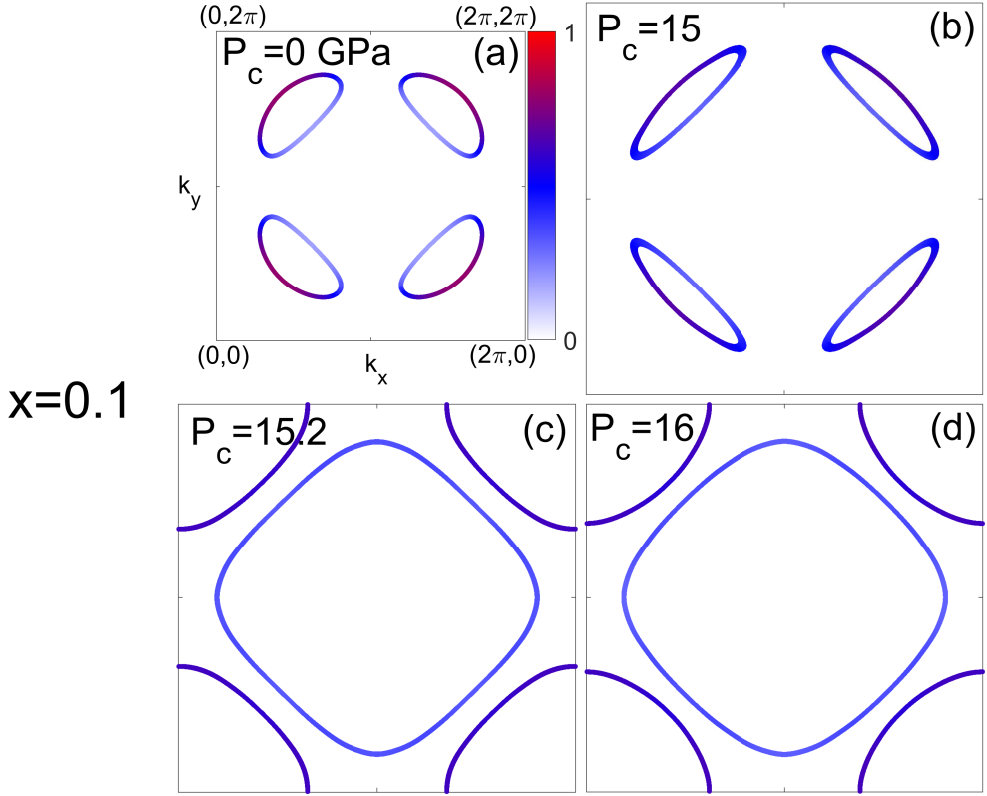

FIG. 8: (Color online) The Fermi contour for the doping $x = 0.1$ at different pressures.

## VI. CONCLUSION

In this work, the electronic structure of quasiparticle excitations in a layer of $CuO_6$ octahedra is studied within the framework of the effective five-band Hubbard model at different values of uniaxial pressure along the $c$ axis and different hole concentrations. This model takes into account quasiparticles both with the origin of the $b_{1g}$ symmetry orbitals arising with the participation of the local Zhang-Rice singlet, and with the character of the $a_{1g}$ symmetry orbitals which are formed by the local two-particle copper-oxygen excited triplet and singlet states. The energy of local electronic states of the $a_{1g}$ symmetry and quasiparticle excitations having the character of these orbitals increases under the influence of uniaxial compression along the $c$ axis. The bands of quasiparticle excitations formed with the participation of the excited triplet and singlet states gradually rises to the top of the valence band; a significant redistribution of contributions from various orbitals occurs in certain $k$-space regions. The contribution of the $p_z$-orbital of apical oxygen increases significantly in the states around the points $(\pi, 0)$, $(0, \pi)$, $(\pi, 2\pi)$, $(2\pi, \pi)$ which form the top of the valence band at high pressures. The crossover of the two-hole local copper-oxygen singlet and triplet states occurs at $P_c = 15.1$ GPa. The character of the transformations of the electronic structure as a result of this crossover is similar to the changes observed with increasing uniaxial compression before the crossover; the degree of these changes increases abruptly due to a sharp increase in the spectral weight of the excitations formed by the Emery-Reiter triplet. A sharp increase in the spectral weight leads to an abrupt increase in the hybridization of quasiparticles and, as the result, the pushing of states in the vicinity of points $(\pi, 0)$,$(0, \pi)$,$(2\pi, \pi)$,$(\pi, 2\pi)$ to the top of the valence band. The change of the first removal state is clearly visible in the picture of how the dispersion of the band just above the Fermi level in the direction $(\pi, 0) - (0, \pi)$ turns inside out, maxima at the points $(\pi/2, \pi/2)$, $(3\pi/2, \pi/2)$, $(\pi/2, 3\pi/2)$, $(3\pi/2, 3\pi/2)$ at $P_c < 15.1$ GPa turn into minima at $P_c > 15.1$ GPa and states at the points $(\pi, 0)$, $(0, \pi)$, $(2\pi, \pi)$, $(\pi, 2\pi)$ become maxima. Uniaxial pressure leads to a transformation of the Fermi contour from small hole pockets around $(\pi/2, \pi/2)$, $(3\pi/2, \pi/2)$, $(\pi/2, 3\pi/2)$, $(3\pi/2, 3\pi/2)$ or large hole and electron contours around $(\pi, \pi)$ to the form of pockets around the points $(0, 0)$, $(0, 2\pi)$, $(2\pi, 0)$, $(2\pi, 2\pi)$ and a large contour around $(\pi, \pi)$ with some particular features for a specific doping.

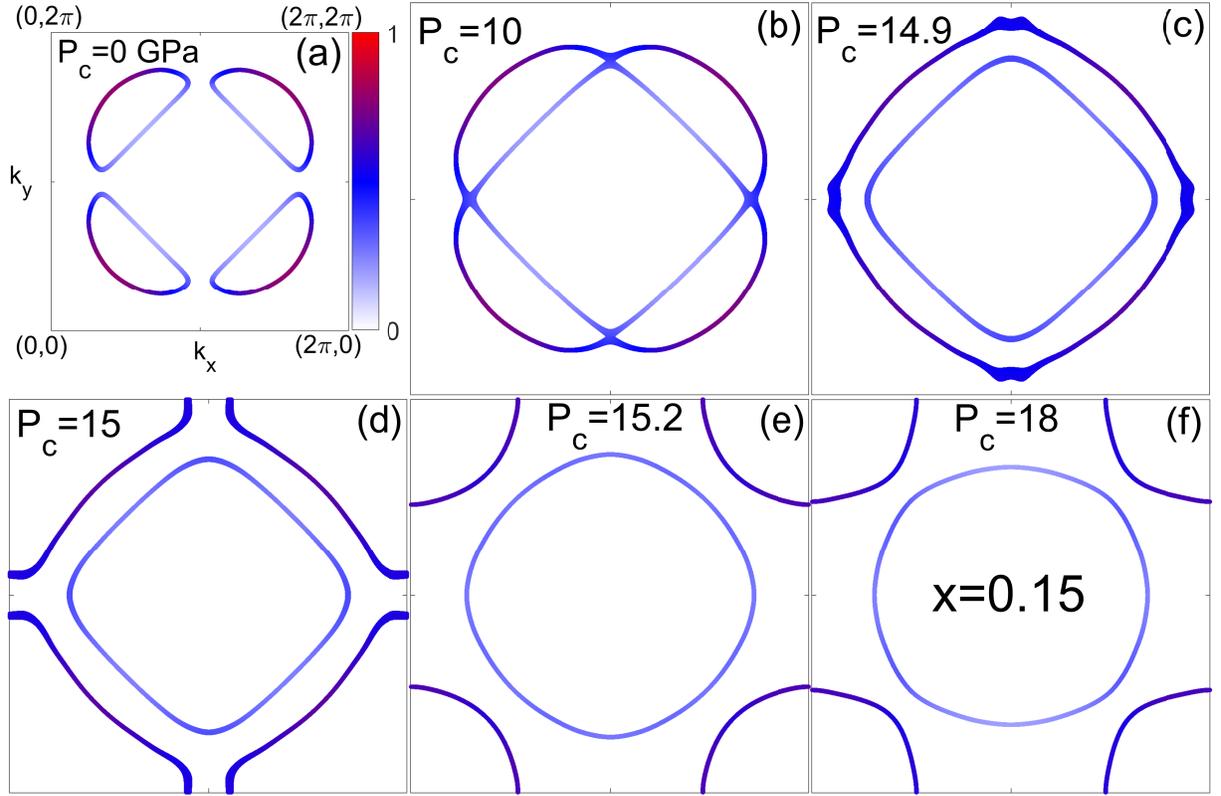

FIG. 9: (Color online) The Fermi contour for the doping $x = 0.15$ at different pressures.

**Acknowledgments**

**Appendix A: The dependences of interatomic distances on pressure**

The pressure for a system in the tetragonal phase is related to the lattice deformation by Hooke's law using elastic constants:

$$
\begin{bmatrix} \sigma_{xx} \\ \sigma_{yy} \\ \sigma_{zz} \\ \sigma_{yz} \\ \sigma_{xz} \\ \sigma_{xy} \end{bmatrix} = \hat{C}\hat{\varepsilon} = \begin{bmatrix} c_{11} & c_{12} & c_{13} & 0 & 0 & 0 \\ c_{12} & c_{22} & c_{23} & 0 & 0 & 0 \\ c_{13} & c_{23} & c_{33} & 0 & 0 & 0 \\ 0 & 0 & 0 & c_{44} & 0 & 0 \\ 0 & 0 & 0 & 0 & c_{55} & 0 \\ 0 & 0 & 0 & 0 & 0 & c_{66} \end{bmatrix} \begin{bmatrix} \varepsilon_{xx} \\ \varepsilon_{yy} \\ \varepsilon_{zz} \\ \varepsilon_{yz} \\ \varepsilon_{xz} \\ \varepsilon_{xy} \end{bmatrix}
$$

(A1)

where $\sigma$ is stress (in GPa), $\varepsilon$ is strain. On the other hand, it can be written:

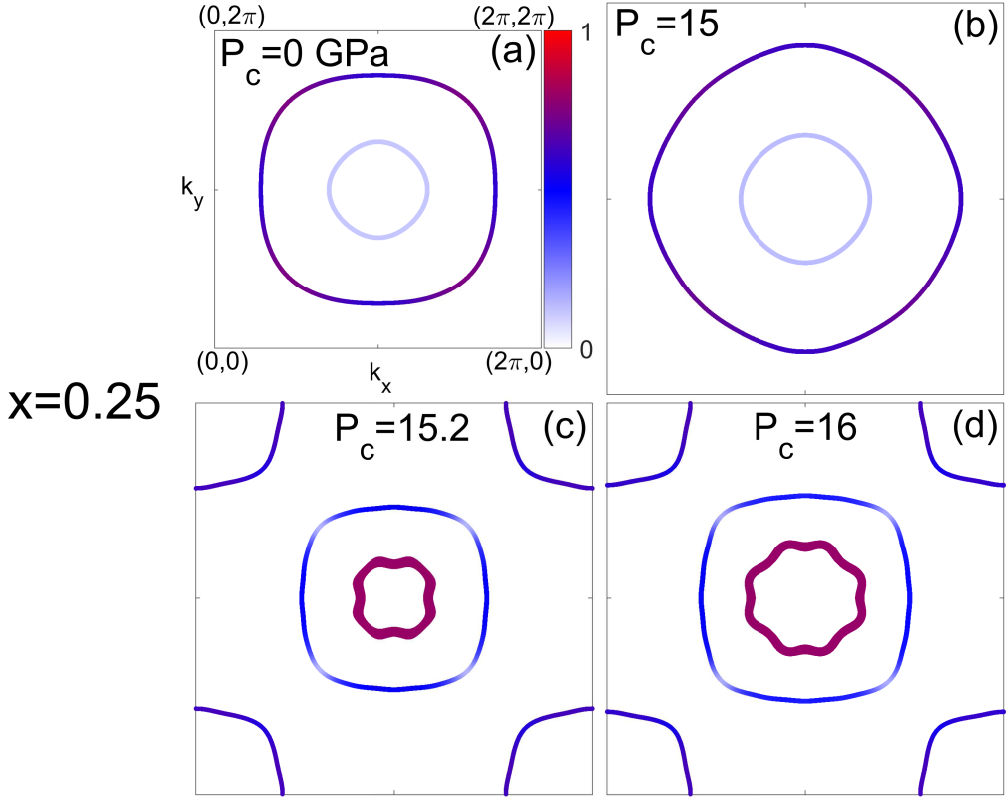

x=0.25

FIG. 10: (Color online) The Fermi contour for the doping $x = 0.25$ at different pressures.

$$
\begin{bmatrix} \varepsilon_1 \\ \varepsilon_2 \\ \varepsilon_3 \\ \varepsilon_4 \\ \varepsilon_5 \\ \varepsilon_6 \end{bmatrix} = \hat{S}\hat{\sigma} = \begin{bmatrix} \frac{1}{E_x} & -\nu_{xy}\frac{1}{E_y} & -\nu_{xz}\frac{1}{E_z} & 0 & 0 & 0 \\ -\nu_{yx}\frac{1}{E_x} & \frac{1}{E_y} & -\nu_{yz}\frac{1}{E_z} & 0 & 0 & 0 \\ -\nu_{zx}\frac{1}{E_x} & -\nu_{zy}\frac{1}{E_y} & \frac{1}{E_z} & 0 & 0 & 0 \\ 0 & 0 & 0 & \frac{1}{G_{yz}} & 0 & 0 \\ 0 & 0 & 0 & 0 & \frac{1}{G_{xz}} & 0 \\ 0 & 0 & 0 & 0 & 0 & \frac{1}{G_{xy}} \end{bmatrix} \begin{bmatrix} \sigma_1 \\ \sigma_2 \\ \sigma_3 \\ \sigma_4 \\ \sigma_5 \\ \sigma_6 \end{bmatrix} \tag{A2}
$$

where $E_i$ are the components of the elastic modulus (Young's modulus, in GPa), $G_{ij}$ are the components of the shear modulus, and $\nu_{ij}$ is the Poisson's ratio which corresponds to contraction along the $i$ axis when pressure is applied along the $j$ axis. The elastic deformations have the form:

$$
\varepsilon_{xx} = \frac{\Delta a}{a_0} = \frac{1}{E_x}\sigma_{xx} - \nu_{xy}\frac{1}{E_y}\sigma_{yy} - \nu_{xz}\frac{1}{E_z}\sigma_{zz} \quad \text{(A3)}
$$

$$
\varepsilon_{yy} = \frac{\Delta b}{b_0} = \nu_{yx}\frac{1}{E_x}\sigma_{xx} + \frac{1}{E_x}\sigma_{yy} - \nu_{yz}\frac{1}{E_z}\sigma_{zz}
$$

$$
\varepsilon_{zz} = \frac{\Delta c}{c_0} = -\nu_{zx}\frac{1}{E_x}\sigma_{xx} - \nu_{zy}\frac{1}{E_y}\sigma_{yy} + \frac{1}{E_z}\sigma_{zz}
$$

Because the stresses $\sigma_{xx} = 0$, $\sigma_{yy} = 0$ for uniaxial pressure along the $c$ axis the elastic deformations take the

form:

$$
\varepsilon_{xx} = \frac{\Delta a}{a_0} = -\nu_{xz}\frac{1}{E_z}\sigma_{zz} \tag{A4}
$$

$$
\varepsilon_{yy} = \frac{\Delta b}{b_0} = -\nu_{yz}\frac{1}{E_z}\sigma_{zz}
$$

$$
\varepsilon_{zz} = \frac{\Delta c}{c_0} = \frac{1}{E_z}\sigma_{zz}
$$

The Young's modulus and the Poisson's ratios can be found if the elastic constants $c_{ij}$ are known. It is necessary to compare matrix elements of $\hat{S}$ with matrix elements of $\hat{C}^{-1}$. Elastic constants for the tetragonal cuprate $La_{1.86}Sr_{0.14}CuO_4$ were found in [67]. The Young's modulus $E_z = 176.45$ GPa and the Poisson's ratio $\nu_{xz} = \nu_{yz} = 0.279$ were obtained on the base of these elastic constants. The lattice parameter $c$ decreases and the lattice parameters $a$ and $b$ increase under compression $\sigma_{zz} < 0$. The decrease in the lattice parameter $c$ is determined by the elastic constant $c_{33}$ ($c_{zzzz}$):

$$\frac{\Delta c}{c_0} = \frac{c_f - c_0}{c_0} = \frac{\sigma_{zz}}{E_z} = -\frac{P}{E_z} \qquad (A5)$$

The increase in the lattice parameter $a$ is determined by the elastic constant $c_{13}$ ($c_{xxzz}$):

$$\frac{\Delta a}{a_0} = \frac{a_f - a_0}{a_0} = -\nu_{xz}\frac{\sigma_{zz}}{E_z} = \nu_{xz}\frac{P}{E_z} \qquad (A6)$$

**Appendix B: The dependences of energy parameters on interatomic distances**

The change in the on-site energy $\varepsilon_{dx}$ under pressure is not considered since the energies $\varepsilon_{dz}, \varepsilon_p, \varepsilon_{pz}$ will be determined as the difference with $\varepsilon_{dx}$. The change in the differences $\varepsilon_p - \varepsilon_{dx}$ and $\varepsilon_{pz} - \varepsilon_p$ with a change in the distance between atoms under uniaxial pressure is taken from the work [54]. In this work, an increase of $\varepsilon_p$ by 2.7% and $\varepsilon_{pz}$ decreased by 19% were obtained under 3% compression along the $c$ axis using $ab$ $initio$ calculations. If we assume a linear variation of $\varepsilon_p$ and $\varepsilon_{pz}$ with pressure then it is possible to obtain their values in a wide pressure range.

The change in the on-site energy $\varepsilon_{dz}$ (or rather, the difference $\varepsilon_{dz} - \varepsilon_{dx}$) with a change in the distance between copper atoms and apical oxygen in the CuO$_6$ octahedron was obtained from simplified calculations of the crystal field in the work [68] based on [69]. The relation of the distance between copper and apical oxygen and the distance between copper and planar oxygen with pressure has the form:

$$\frac{r_{Cu-O_{ap}}}{r_{Cu-O_{pl}}} = \frac{c_{ap0} - \Delta c_{ap}}{\frac{a_0 + \Delta a}{2}} = \frac{r^0_{Cu-O_{ap}}}{r^0_{Cu-O_{pl}}}\frac{(E_z + P)}{(E_z + \nu_{xz}P)}. \qquad (B1)$$

The ratio of these distances between copper and apical oxygen and between copper and planar oxygen for the undeformed La$_{2-x}$Sr$_x$CuO$_4$ compound is $\frac{r^0_{Cu-O_{ap}}}{r^0_{Cu-O_{pl}}} = 1.2748$, lattice parameters and distances between atoms are taken for the undeformed La$_{1.9}$Sr$_{0.1}$CuO$_4$ from [70].

Since we need to define the dependence of the parameter $t_{dzpz}$ on pressure it is necessary to know not the change in the lattice parameter $c$ with pressure but the change in the distance between the copper atom and the apical oxygen atom with pressure. The change in both the lattice parameter $c$ and the distance between the copper atom and the apical oxygen atom $r_{Cu-O_{ap}}$ in the Hg1201 compound was studied in the work [71]. It is clear from the results of this work that the relative change $\Delta c/c_0$ in the lattice parameter $c$ under pressure is less than the change in the distance between the copper atom and the apical oxygen atom $\Delta c_{ap}/c_{ap}$. Thus, the relative change $\Delta c/c_0$ can be considered as a lower bound for the relative change $\Delta c_{ap}/c_{ap}$, the actual change $\Delta c_{ap}/c_{ap}$ with pressure may be even more significant. The compounds Hg1201 and La$_{2-x}$Sr$_x$CuO$_4$ have similar structures and the same basic structural elements therefore it can be assumed that La$_{2-x}$Sr$_x$CuO$_4$ will have a similar behavior of lattice parameters with pressure.

The hopping integrals between copper and oxygen orbitals are proportional to the interatomic distance with the power of $-7/2$ [72]. That is

$$t_{dzpz} \sim \frac{1}{r^{7/2}_{Cu-O_{ap}}} = \frac{1}{(c_{ap0} + \Delta c_{ap})^{7/2}} = \frac{1}{c^{7/2}_{ap0}\left(1 - \frac{P}{E_z}\right)^{7/2}} \qquad (B2)$$

The ratios of the hopping integrals $t_{dzpz}$ in the undeformed and deformed states:

$$t_{dzpz}(c_{ap0} + \Delta c_{ap}) = \frac{E_z^{7/2}}{(E_z - P)^{7/2}}t_{dzpz}(c_{ap0}) \qquad (B3)$$

The ratios of the hopping integrals $t_{pd}$ and $t_{pdz}$ in the undeformed and deformed states:

$$t_{pd}\left(\frac{a_0 + \Delta a}{2}\right) = \frac{E_z^{7/2}}{(E_z + \nu_{xz}P)^{7/2}}t_{pd}\left(\frac{a_0}{2}\right), \qquad (B4)$$

$$t_{pdz}\left(\frac{a_0 + \Delta a}{2}\right) = \frac{E_z^{7/2}}{(E_z + \nu_{xz}P)^{7/2}}t_{pdz}\left(\frac{a_0}{2}\right)$$

The hopping integrals between oxygen $p$-orbitals are proportional to the power $-2$ of the interatomic distance [72]. Then the ratios of the hopping integrals $t_{pp}$ and $t_{ppz}$ in the undeformed and deformed states are determined as follows:

$$t_{pp}\left(a/\sqrt{2}\right) = \frac{E_z^2}{(E_z + \nu_{xz}P)^2} t_{pp}\left(a_0/\sqrt{2}\right), \tag{B5}$$

$$t_{ppz}\left(\frac{1}{2}\sqrt{a^2+c_{ap}^2}\right) = t_{ppz0}\left(\frac{1}{2}\sqrt{a_0^2+c_{ap0}^2}\right) \frac{\left(\sqrt{a_0^2+c_{ap0}^2}\right)^2}{\left(a_0\sqrt{\left(1+\frac{c_{ap0}^2}{a_0^2}+2\left(\nu_{xz}-\frac{c_{ap0}^2}{a_0^2}\right)\left(\frac{P}{E_z}\right)+\left(\nu_{xz}^2+\frac{c_{ap0}^2}{a_0^2}\right)\left(\frac{P}{E_z}\right)^2\right)}\right)^2}$$

[1] G. Hearne, M. Pasternak, R. Taylor, and P. Lacorre, Phys. Rev. B **51**, 11495 (1995).

[2] W. Xu, O. Naaman, G. Rozenberg, M. Pasternak, and R. Taylor, Phys. Rev. B **64**, 094411 (2001).

[3] I. Troyan, A. Gavriliyuk, V. Sarkisyan, I. Lyubutin, R. Rüffer, O. Leupold, A. Barla, B. Doyle, and A. Chumakov, JETP Lett. **74**, 24 (2001).

[4] V. Sarkisyan, I. Troyan, I. Lyubutin, A. Gavriyuk, and A. Kashuba, JETP Lett. **76**, 664 (2002).

[5] A. Gavriliuk, I. Trojan, S. Ovchinnikov, I. Lyubutin, and V. Sarkisyan, J. Exp. Theor. Phys. **99**, 566 (2004).

[6] I. Lyubutin, A. Gavriliuk, V. Struzhkin, S. Ovchinnikov, S. Kharlamova, L. Bezmaternykh, M. Hu, and P. Chow, JETP Lett. **84**, 518 (2007).

[7] I. Lyubutin, A. Gavriliuk, I. Trojan, and R. Sadykov, JETP Lett. **82**, 702 (2005).

[8] I. Lyubutin, A. Gavriliuk, and V. Struzhkin, JETP Lett. **88**, 524 (2008).

[9] M. Pasternak, G. Rozenberg, G. Machavariani, R. T. O. Naaman, and R. Jeanloz, Phys. Rev. Lett. **82**, 4663 (1999).

[10] I. Lyubutin and A. Gavriliuk, Bull. Russ. Acad. Sci. Phys. **71**, 1594 (2007).

[11] I. Lyubutin, A. Gavriliuk, and V. Struzhkin, MRS Symp. Proc. **987**, 167 (2007).

[12] R. Cohen, I. Mazin, and D. Isaak, Science **275**, 654 (1997).

[13] Y. Orlov, S. Nikolaev, V. Dudnikov, V. Gavrichkov, and S. Ovchinnikov, Phys.Usp. **66**, 647 (2023).

[14] Z. Fang, I. Solovyev, H. Sawada, and K. Terakura, Phys. Rev. B **59**, 762 (1999).

[15] D. Kasinathan, J. Kuneš, K. Koepernik, C. Diaconu, R. Martin, I. Prodan, G. Scuseria, N. Spaldin, L. Petit, T. Schulthess, et al., Phys. Rev. B **74**, 195110 (2006).

[16] D. Kasinathan, K. Koepernik, and W. Pickett, New J. Phys. **9**, 235 (2007).

[17] J.-G. Cheng, K. Matsubayashi, W. Wu, J. Sun, F. Lin, J. Luo, and Y. Uwatoko, Phys. Rev. Lett. **114**, 117001 (2015).

[18] Y. Wang, L. Bai, T. Wen, L. Yang, H. Gou, Y. Xiao, P. Chow, M. Pravica, W. Yang, and Y. Zhao, Angewandte Chemie **128**, 10506 (2016).

[19] A. Dyachenko, A. Lukoyanov, A. Shorikov, and V. Anisimov, Phys. Rev. B **98**, 085139 (2018).

[20] S. Kimber, A. Salamat, S. Evans, H. Jeschke, K. Muthukumard, M. Tomi, F. Salvat-Pujold, R. Valentíd, M. Kaisheva, I. Zizakf, et al., Proc. Natl. Acad. Sci. USA **111**, 5105 (2014).

[21] T. Hung, C. Huang, L. Deng, M. Ou, Y. Chen, M. Wu, S. Huyan, C. Chu, P. Chen, and T. Lee, Nat. Commun. **12**, 5436 (2021).

[22] M. Yi, Y. Zhang, Z.-X. Shen, and D. Lu, npj Quant. Mater. **2**, 57 (2017).

[23] J.-H. Chu, J. Analytis, K. D. Greve, P. McMahon, Z. Islam, Y. Yamamoto, and I. Fisher, Science **329**, 824 (2010).

[24] A. Chubukov, Annu. Rev. Condens. Matter Phys. **3**, 57 (2012).

[25] R. Fernandes, A. Chubukov, and J. Schmalian, Nature Phys. **10**, 97 (2014).

[26] F.-C. Hsu, J.-Y. Luo, K.-W. Yeh, T.-K. Chen, T.-W. Huang, P. M. Wu, Y.-C. Lee, Y.-L. Huang, Y.-Y. Chu, D.-C. Yan, et al., Proc. Natl. Acad. Sci. USA **105**, 14262 (2008).

[27] Y. Mizuguchi, F. Tomioka, S. Tsuda, T. Yamaguchi, and Y. Takano, Appl. Phys. Lett. **93**, 152505 (2008).

[28] Y. Mizuguchi and Y. Takano, J. Phys. Soc. Jpn. **79**, 102001 (2010).

[29] Y. Kamihara, T. Watanabe, M. Hirano, and H. Hosono, J. Am. Chem. Soc. **130**, 3296 (2008).

[30] M. Rotter, M. Tegel, and D. Johrendt, Phys. Rev. Lett. **101**, 107006 (2008).

[31] M. Wu, F. Hsu, K. Yeh, T. Huang, J. Luo, M. Wang, H. Chang, T. Chen, S. Rao, B. Mok, et al., Physica C **469**, 340 (2009).

[32] D. C. Johnston, Adv. Phys. **59**, 803 (2010).

[33] G. Stewart, Rev. Mod. Phys. **83**, 1589 (2011).

[34] M.-K. Wu, M.-J. Wang, and K.-W. Yeh, Sci. Technol. Adv. Mater. **14**, 014402 (2013).

[35] K. Deguchi, Y. Takano, and Y. Mizuguchi, Sci. Technol. Adv. Mater. **13**, 054303 (2012).

[36] E. Dagotto, Rev. Mod. Phys. **85**, 849 (2013).

[37] C. de la Cruz, Q. Huang, J. Lynn, J. Li, W. R. II, J. Zarestky, H. Mook, G. Chen, J. Luo, N. Wang, et al., Nature **453**, 899 (2008).

[38] P. Dai, Rev. Mod. Phys. **87**, 855 (2015).

[39] H. Luo, R. Zhang, M. Laver, Z. Yamani, M. Wang, X. Lu, M. Wang, Y. Chen, S. Li, S. Chang, et al., Phys. Rev. Lett. **108**, 247002 (2012).

[40] S. Margadonna, Y. Takabayashi, M. McDonald, K. Kasperkiewicz, Y. Mizuguchi, Y. Takano, A. Fitch, E. Suarde, and K. Prassides, Chem. Commun. **43**, 5607 (2008).

[41] S. Medvedev, T. McQueen, I. Troyan, T. Palasyuk, M. Eremets, R. Cava, S. Naghavi, F. Casper, V. Ksenofontov, G. Wortmann, et al., Nature Mater. **8**, 630

(2009).

[42] S. Margadonna, Y. Takabayashi, Y. Ohishi, Y. Mizuguchi, Y. Takano, T. Kagayama, T. Nakagawa, M. Takata, and K. Prassides, Phys. Rev. B **80**, 064506 (2009).

[43] T. Imai, K. Ahilan, F. Ning, T. McQueen, and R. Cava, Phys. Rev. Lett. **102**, 177005 (2009).

[44] M. Bendele, A. Amato, K. Conder, M. Elender, H. Keller, H.-H. Klauss, H. Luetkens, E. Pomjakushina, A. Raselli, and R. Khasanov, Phys. Rev. Lett. **104**, 087003 (2010).

[45] M. Bendele, A. Ichsanow, Y. Pashkevich, L. Keller, T. Strässle, A. Gusev, E. Pomjakushina, K. Conder, R. Khasanov, and H. Keller, Phys. Rev. B **85**, 064517 (2012).

[46] F. Zhang and T. Rice, Phys. Rev. B **37**, 3759 (1988).

[47] V. Emery and G. Reiter, Phys. Rev. B **38**, 11938 (1988).

[48] C. Meingast, O. Kraut, T. Wolf, H. Wühl, A. Erb, and G. Müller-Vogt, Phys. Rev. Lett. **67**, 1634 (1991).

[49] F. Hardy, N. Hillier, C. Meingast, D. Colson, Y. Li, N. Barišić, G. Yu, X. Zhao, M. Greven, and J. Schilling, Phys. Rev. Lett. **105**, 167002 (2010).

[50] F. Gugenberger, C. Meingast, G. Roth, K. Grube, V. Breit, T. Weber, H. Wühl, S. Uchida, and Y. Nakamura, Phys. Rev. B **49**, 13137 (1994).

[51] C. Meingast, A. Junod, and E. Walker, Physica C **272**, 106 (1996).

[52] J. Schilling and S. Klotz (1992).

[53] V. Gavrichkov, S. Ovchinnikov, and G. Ulm, Fiz. Tverd. Tela **49**, 580 (2007).

[54] K. Sidorov, V. Gavrichkov, S. Nikolaev, Z. Pchelkina, and S. Ovchinnikov, Phys. Status Solidi B **253**, 486–(2016).

[55] S. Ovchinnikov and I. Sandalov, Physica C **161**, 607 (1989).

[56] V. Gavrichkov, S. Ovchinnikov, A. Borisov, and E. Goryachev, J. Exp. Theor. Phys. **91**, 369 (2000).

[57] S. Ovchinnikov and V. Val'kov, *Hubbard operators in the Theory of Strongly correlated electrons* (Imperial College Press, London-Singapore, 2004).

[58] M. Korshunov, V. Gavrichkov, S. Ovchinnikov, I. Nekrasov, Z. Pchelkina, and V. Anisimov, Phys. Rev. B **72**, 165104 (2005).

[59] I. Makarov, V. Gavrichkov, E. Shneyder, I. Nekrasov, A. Slobodchikov, S. Ovchinnikov, and A. Bianconi, J. Supercond. Nov. Magn. **32**, 1927 (2019).

[60] B. Shastry, Phys. Rev. Lett. **63**, 1288 (1989).

[61] R. Raimondi, J. Jefferson, and L. Feiner, Phys. Rev. B **53**, 8774 (1996).

[62] N. Plakida, L. Anton, S. Adam, and G. Adam, J. Exp. Theor. Phys. **97**, 331 (2003).

[63] I. Makarov and S. Ovchinnikov, Phys. Lett. A **444**, 128226 (2022).

[64] V. Val'kov and D. Dzebisashvili, J. Exp. Theor. Phys. **100**, 608 (2005).

[65] M. Korshunov and S. Ovchinnikov, Eur. Phys. J. B **57**, 271 (2007).

[66] V. Gavrichkov and S. Ovchinnikov, Fiz. Tverd. Tela **50**, 1037 (2008).

[67] A. Migliori, W. Visscher, S. Wong, S. Brown, I. Tanaka, H. Kojima, and P. Allen, Phys. Rev. Lett. **64**, 2458 (1990).

[68] R. Fumagalli, A. Nag, S. Agrestini, M. Garcia-Fernandez, A. Walters, D. Betto, N. Brookes, L. Braicovich, K.-J. Zhou, G. Ghiringhelli, et al., Physica C **581**, 1353810 (2021).

[69] I. Bersuker, *Electronic structure and properties of transition metal compounds* (John Wiley & Sons, 2010).

[70] H. Takahashi, H. Shaked, B. Hunter, P. Radaelli, R. Hitterman, D. Hinks, and J. Jorgensen, Phys. Rev. B **50**, 3221– (1994).

[71] S. Wang, J. Zhang, J. Yan, X.-J. Chen, V. Struzhkin, W. Tabis, N. Barišić, M. K. Chan, C. Dorow, X. Zhao, et al., Phys. Rev. B **89**, 024515 (2014).

[72] W. Harrison, *Electronic Structure and the Properties of Solids: The Physics of the Chemical Bond* (W.H. Freeman and Company, San Francisco, 1983).